\documentclass[9pt, twocolumn]{extarticle}

\usepackage{authblk}
\usepackage{amssymb}

\makeatletter
\renewcommand{\maketitle}{\bgroup\setlength{\parindent}{0pt}
\begin{flushleft}
  \textbf{\@title}

  \@author
\end{flushleft}\egroup
}
\makeatother

\title{\Huge 
Fair Money \textendash \;Public Good Value Pricing With Karma Economies
\vspace{0.5cm}}
\author{
\Large 
 Kevin Riehl$^{1,*}$ ,
 Anastasios Kouvelas$^{1}$ , 
 Michail A. Makridis$^{1}$ \\
\tiny 
[*] Corresponding author: kriehl@ethz.ch \\
[1] Traffic Engineering group, Institute for Transport Planning and Systems, ETH Zurich, Stefano-Franscini-Platz 5, Zurich, 8093, Switzerland
}

\date{\today}

\usepackage[showframe=false, margin=.65in]{geometry}

\usepackage{abstract}
\renewcommand{\abstractname}{}    

\usepackage{lipsum}

\usepackage{afterpage}

\renewenvironment{abstract}
 {\small
  \begin{center}
  \bfseries \abstractname\vspace{-.5em}\vspace{0pt}
  \end{center}
  \list{}{%
    \setlength{\leftmargin}{10mm}
    \setlength{\rightmargin}{\leftmargin}%
  }%
  \item\relax}
 {\endlist}

\usepackage[utf8]{inputenc}

\usepackage{wrapfig}
\usepackage[lofdepth,lotdepth]{subfig}
\usepackage{booktabs}

\usepackage{rotating}
\usepackage{caption}

\usepackage{csquotes}

\usepackage{todonotes}

\usepackage{stfloats}

\usepackage[T1]{fontenc}
\usepackage{textcomp}
\usepackage{times}

\usepackage{framed} 

\usepackage[citestyle=nature, 
            maxcitenames=2,
			bibstyle=nature, 
			language=auto,
			url=false,
			backend=bibtex,
			doi=false,
			isbn=false]{biblatex}
	\renewbibmacro*{volume+number+eid}{%
  \printfield{volume}%
  \setunit*{\addnbspace}
  \printfield{number}%
  \setunit{\addcomma\space}%
  \printfield{eid}}
\DeclareFieldFormat[article]{number}{\mkbibparens{#1}}
\usepackage{setspace}
\usepackage{color}
\definecolor{black}{gray}{0} 

\usepackage[colorlinks=true,linkcolor=black,citecolor=black]{hyperref}

\usepackage[]{fancyhdr}
{\renewcommand{\markboth}[2]{}
\fancyhead[R,RO]{}
\fancyhead[L,LO]{}
\pagestyle{fancy}
\fancyhead[RO,R]{
\textit{
\\\hrulefill\\
Fair Money \textendash \; (Riehl et al. 2024) \; \textendash \; A Submission To: npj | sustainable mobility and transport (2024) }
}

\usepackage{algorithm}
\usepackage{algpseudocode}
\newcounter{algsubstate}
\renewcommand{\thealgsubstate}{\alph{algsubstate}}
\newenvironment{algsubstates}
  {\setcounter{algsubstate}{0}%
   \renewcommand{\State}{%
     \stepcounter{algsubstate}%
     \Statex {\footnotesize \; \; \; \; \thealgsubstate:}\space}}
  {}

\usepackage{tabularx}
\newcolumntype{s}{>{\hsize=.5\hsize}X}

\usepackage{ntheorem}

\begin{document}

\twocolumn[
    \begin{@twocolumnfalse}
        \maketitle
        \begin{abstract}
            City road infrastructure is a public good, and over-consumption by self-interested, rational individuals leads to traffic jams.
            Congestion pricing is effective in reducing demand to sustainable levels, but also controversial, as it introduces equity issues and systematically discriminates lower-income groups.            
            Karma is a non-monetary, fair, and efficient resource allocation mechanism, that employs an artificial currency different from money, that incentivizes cooperation amongst selfish individuals, and achieves a balance between giving and taking.
            Where money does not do its job, Karma achieves socially more desirable resource allocations by being aligned with consumers' needs rather than their financial power. 
            This work highlights the value proposition of Karma, gives guidance on important Karma mechanism design elements, and equips the reader with a useful software framework to model Karma economies and predict consumers' behaviour.
            A case study demonstrates the potential of this feasible alternative to money, without the burden of additional fees.
            \hfill \break \;
            \newline
            \textbf{Keywords: Resource Allocation, Fairness, Karma mechanism, Non-monetary markets, Artificial currencies, Public Goods}
            \newline
            \textbf{JEL Classification: B26, B52, C02, C68, C73, C88, D47, D58, D63, H41 }
            
        \end{abstract}
    \end{@twocolumnfalse}
]


\section{Introduction}

\subsection{Background and Motivation}

Public goods are resources available to all in society without restriction.
The tragedy of the commons describes the situation, when self-interest drives individuals to overuse, depletion, or damage these goods (even this is not in anyone's long-term interest)~\cite{hardin1968tragedy,kaul1999global}.
Road networks and congestion exemplify this problem.
Road networks attract many users, as they enable individual mobility, and provide accessibility to many useful destinations for work, shopping, education, socialization, or recreation.
Roads provide a diminishing utility for a growing number of users; 
if too many vehicles enter the network, traffic slows down and congestion arises.
Congestion is a global issue, with consequences for drivers, residents, and society.
Noise, air pollution, and security incidents, affect the living quality and health of residents.
A significant amount of a driver's life-time is wasted in traffic jams.
Wasted consumption of energy and time cause financial damages to the economy, 
and avoidable emissions contribute to global warming and the climate change~\cite{arnott1994economics,koetse2009impact,zhang2013air}. 

Governmental intervention and regulation can help to solve congestion, by aligning individual incentives with the collective good and keeping consumption of the road infrastructure at sustainable levels.
Access-restricting regulations involve the establishment of property rights, rationing (capping)~\cite{gentile2005advanced}, cap-and-trade mechanisms~\cite{schmalensee2017lessons}, taxation~\cite{barnett1980pigouvian}, or value pricing~\cite{small2001value}.
A broad variety of economic instruments have been proposed to cope with the issue of congestion and can be used to control traffic demand and supply.
Traffic demand management employs road pricing, such as tolled bridges and tunnels, urban congestion pricing, and tolled highway lanes~\cite{santos2006road}.
Besides, examples to control the supply can be found in license plate rationing, tradeable credit schemes and mobility permits~\cite{nie2017license,provoost2023design}.
These economic instruments can be designed and implemented in conjunction with other traffic management strategies, such as infrastructure improvements, public transport expansion and investments, traffic signal optimization, and information provision, to create a comprehensive and effective transportation system.

Economic instruments introduce monetary, market-based incentive mechanisms, for the allocation of resources.
Despite potential benefits for solving the socially relevant question of traffic congestion, economic instruments appear to enjoy little support outside academia. 
Limited social and political support has caused many proposed schemes to be abandoned before implementation, or postponed for an undefined time~\cite{lindsney2001traffic}.
Reasons for the lacking public acceptance include equity concerns, lack of trust in government, perceived severity of congestion, and organized opposition by drivers. 
Economic instruments can raise equity concerns; 
they may disproportionately benefit higher-income travelers who can afford tolls, while imposing additional costs on lower-income commuters with fewer transportation alternatives. 
Often, the public is not convinced about the severity of congestion and the need for the implementation of economic instruments, doubts technological feasibility, or is concerned with privacy issues due to a lack of trust into the government~\cite{wang2021support,moeinaddini2024acceptability}.
Moreover, there is refusal by drivers to being charged for something they feel is not their fault and ought to be free to them~\cite{lindsney2001traffic}.

Monetary markets are not always the right tool for resource allocation, and in many contexts the use of money is not desired, socially not accepted, considered unethical, or even prohibited.
Therefore, a growing branch of literature is concerned with artificial currencies~\cite{gorokh2021monetary}, which represent non-monetary markets and resource allocation mechanisms.
For isolated, single-stage resource allocation problems, extensive work on non-monetary matching and combinatorial assignment problems has been conducted~\cite{budish2011combinatorial,dutot2011approximation,skowron2013non,carreras2016theory,nguyen2016assignment,karaenke2020non}.
For repeated resource allocation problems, there are few works on non-monetary market mechanisms yet, among which Karma has evolved as an important narrative~\cite{riehl2024karma}.

Karma employs a currency different from money; it can only be gained by producing and only be lost by consuming a specific resource.
It is a resource-inherent, non-monetary, non-tradeable, artificial currency for prosumer resources (produced and consumed by market participants likewise).
As a non-monetary mechanism, Karma complements monetary markets and provides attractive properties. 
For example, it is fairness-enhancing, near incentive-compatible, and robust towards population heterogeneity~\cite{article_567,article_1002,article_600}. 
Due to its design, one can consider Karma as playing against your future self, as the only way to consume is to put effort and produce first, and future needs must be traded off against present needs when consuming.
Last but not least, Karma is reported to not only concern the efficiency and fairness of resource allocation but also lead to a decrease in resource scarcity in peer-to-peer markets~\cite{article_foodbank,article_organ_A,article_organ_B,article_organ_C}.

Let us discuss Karma at the example of congestion pricing~\cite{de2011traffic}, where vehicles need to pay a surcharge for driving in the city. 
Due to the monetary disincentive, people will trade off their urgency to drive to the city, and their willingness to pay.
However, congestion pricing can be problematic, as equity issues emerge in a society with unequal distribution of economic power: the poorest will most likely not be able to afford the charge, and thus consume significantly less.
Karma could make a difference here: not driving could be considered as producing, and driving could be considered as consuming the resource ``right of driving to the city''. 
Instead of paying money as a congestion pricing tax, Karma points could be used. 
These Karma points cannot be bought, but only gained by not consuming.
Therefore, Karma would create a balance between giving and taking, between using and not using the public good.
This would force individuals not to trade off the price with other resources they could buy alternatively but to solely consider present versus future consumption of this specific resource.
Moreover, the socio-economic contexts, such as income or wealth, and therefore the above-mentioned equity considerations, would not play a role anymore.
Finally, contrary to a monetary pricing, Karma would not impose additional financial costs to the society.
 
\subsection{Objectives}

The overall goal of this study is to demonstrate the potential of Karma to address the equity issues of economic instruments when coping with public goods.
This work follows mainly two goals:
(i) highlight the value proposition of Karma as a non-monetary resource allocation mechanism, 
(ii) equip the reader with the necessary tools and knowledge to successfully apply Karma in various contexts and domains.
To achieve these objectives, we elaborate on the properties of Karma, provide guidance on the design of Karma mechanisms,
outline a game theoretic model of Karma as a dynamic population game, and present a software framework to model problems and predict user behaviour in Karma economies.
In addition, we compare Karma economies over monetary markets in a case study on bridge tolling, to demonstrate its usefulness.

This study contributes to the economic discussion of traffic demand management by providing the perspective of non-monetary market mechanisms to cope with equity issues.
Furthermore, this study contributes to the literature on Karma mechanisms, as it provides a unifying framework of mechanism design elements, as well as a software library to efficiently simulate Karma economies. Ultimately, it enables more systematic research on Karma for resource allocation.

The brief remainder of this work is as follows.
Section~\ref{sec:litreview} reviews related works on Karma, presents applications in transportation, highlights its value proposition, and presents mechanism design elements.
Section~\ref{sec:methods} outlines the modelling of Karma as a game, presents the Software Framework and outlines its usage on a computation example.
Section~\ref{sec:casestudy} presents the case study of a tolled bridge and the underlying assumptions we have made when comparing money with Karma.
Section~\ref{sec:results} presents and discusses the results of comparing money and Karma markets.
Section~\ref{sec:conclusions} concludes this work and outlines future research directions.


\section{Literature Review}  \label{sec:litreview}


\subsection{Related Works on Karma}

Karma is a concept that emerged from the domain of filesharing~\cite{vishnumurthy2003karma}, enjoyed popularity as a technological component to blockchain applications, and gained prominence as an artificial, non-monetary currency in the literature of economics.
As a resource allocation mechanism, Karma was applied in a wide range of applications, including file and computational resource sharing in peer-to-peer networks, allocation of transmission bandwidth in telecommunication networks, and distribution of food and organ donations~\cite{riehl2024karma}.

In the context of traffic demand management, pilot studies on Karma investigated its potential use for high occupancy and priority toll lanes~\cite{article_1002,article_1004}, auction-controlled intersection management with fully connected vehicles~\cite{article_567,article_600}, and transportation modality pricing~\cite{article_265,article_z02}.


When applied as a resource allocation mechanism, Karma can be described by a population of agents, where each agent $\cdots$

\begin{itemize}
    \setlength\itemsep{0em}
    \item has a specific amount of Karma
    \item has a random, time-varying urgency (represents the agent's cost when not getting a specific resource)
    \item has an individual temporal consumption preference type (discount factor, which represents the subjective trade-off between consuming now versus later)
\end{itemize}

In a repeated, auction-like setup, agents are matched randomly in rounds to compete for a specific resource, by bidding with Karma.
Depending on their urgency, Karma balance, and consumption type, agents must determine an optimal bid to earn the resource when necessary, while accounting for potential future competitions in following rounds.

\begin{table}[h]
    \caption{\textbf{Karma mechanism design elements}}
    \label{table_design_parameters}
    \begin{tabular}{@{}ll@{}}
        \toprule
        Design Element & Option \\
        \midrule
        \textbf{Currency}  & \\
            \hspace*{0.3cm} Parity         & Price, Threshold, Binary \\
            \hspace*{0.3cm} Balance limits & Unlimited, Bounded (upper, lower) \\
            \hspace*{0.3cm} Amount control & \parbox[t][][t]{5cm}{Constant, Constant per capita, Uncontrolled, Expiry} \\
            \hspace*{0.3cm} Initialization & \parbox[t][][t]{5cm}{Equal endowment, Weighted endowment, Random endowment, None} \\
            \hspace*{0.3cm} Redistribution & \parbox[t][][t]{5cm}{Property tax, Payment tax, Lottery, Expiry, None} \\
         & \\
        \textbf{Interaction}  & \\
            \hspace*{0.3cm} Price control & Auction, Centrally defined, None \\
            \hspace*{0.3cm} Price limits  & Only positive, Fix , None \\
            \hspace*{0.3cm} Resource provision & By one agent, By all agents, By system \\
            \hspace*{0.3cm} Resource allocation & \parbox[t][][t]{5cm}{Auction winner, System decision, Provider decision} \\
            \hspace*{0.3cm} Counter-party & N agents, One agent, System \\
            \hspace*{0.3cm} Peer selection & \parbox[t][][t]{5cm}{Market, Neighbourhood, Randomly assigned, Active selection} \\
            \hspace*{0.3cm} Decision-making & Free \\
            \hspace*{0.3cm} Urgency process & Homogeneous, Heterogeneous \\
            \hspace*{0.3cm} Temporal preference & Homogeneous, Heterogeneous \\
         & \\
        \textbf{Transaction}  & \\
            \hspace*{0.3cm} Payment amount & \parbox[t][][t]{5cm}{Bid, Peer's bid, Difference in bids, Fixed, Nothing} \\
            \hspace*{0.3cm} Payment receiver & \parbox[t][][t]{5cm}{Resource provider, System, Equally across population, Weighted across population} \\
            \hspace*{0.3cm} Karma gain & Resource provision, Resource consumption  \\
            \hspace*{0.3cm} Karma loose & \parbox[t][][t]{5cm}{Resource consumption, Expiration, Rule-violation}  \\
         & \\
        \midrule
    \end{tabular}
\end{table}

\cite{riehl2024karma} identifies mechanism design elements based on a systematic comparison of previous Karma applications. These mechanism design elements are outlined in Table~\ref{table_design_parameters}, and cover three aspects of Karma mechanisms: currency, interaction, and transaction.
As it turned out in many of the Karma applying works, a major design complexity is choosing the right amount of Karma currency in circulation~\cite{article_scrips,article_185,article_265,article_z02}. 
If there are too few currency units, there will be hoarding to save the scarce currency for very urgent situations to consume; 
if there are too many currency units, the value of a single value is not sufficient anymore to stimulate provision of resources.
In case of a time-variant resource supply, a dedicated amount control becomes necessary.

\subsection{Value proposition of Karma}

Karma has the potential to address the equity issues of economic instruments when coping with public goods.
Governmental intervention to cope with public goods is regulating the access to it, so that consumption is limited to a sustainable amount.
The access right to consuming the public good can be considered as the resource of interest.
This resource can be considered as a prosumer resource, as each system participant can produce and consume it.
The allocation of this resource can be done by a central coordinator (i.e.\ the government) or a decentralized mechanism (i.e.\ the market).
Monetary markets can cause equity issues, as resource consumption is linked to economic power, which is usually distributed unequally.
Instead of monetary markets, the use of the artificial currency mechanisms such as Karma could be used when allocating resources.

Previous research has found that Karma provides useful features that distinguish it from money.
\textbf{Karma is fairness-enhancing}, as the consumption of resources depends on the urgency and previous behaviour, and not on economic power.
Karma is able to approach levels of efficiency similar to centralized, efficiency-maximizing algorithms, while outperforming in terms of fairness~\cite{article_567,article_600,article_1005}.
\textbf{Karma has a direct approach towards utility}. 
As Karma is resource-inherent, there is no other aspect besides the pure utility value of a specific resource for participants when bidding at auctions. 
Hence, Karma achieves high levels of incentive compatibility~\cite{krishna2009auction,moulin1980strategy,article_243,article_600}.
In monetary markets, the readiness to pay prices not only depends on utility, but also on economic power, and the comparison of values with other resources that could be bought alternatively for this price.
Karma is of substantial value here, as it enables an intuitive, direct, utility-focused, comparison-free evaluation of a resource's value.
Rather than to trade-off resources against resources as in monetary mechanisms, Karma allows only the resource-specific trade-off between present and future needs.
Besides, \textbf{Karma decreases the scarcity of resources}, as it incentivizes cooperative behaviour and contributions amongst a population of rational, selfish individuals. 
This happens, when resources are not provided by a central coordinator, but provided by prosumers themselves (e.g.\ services). 
The underlying incentive scheme of Karma can cause significant increases of available resources; examples for this property include content sharing~\cite{vishnumurthy2003karma,vishnumurthy2008substrate,article_243}, computation power in distributed computing applications~\cite{vishnumurthy2003karma,vishnumurthy2008substrate,article_243,article_492}, a better mobile network coverage~\cite{article_185,article_273,article_348,article_55,concept_nuglets1,concept_nuglets2}, and more food and organ donations~\cite{article_foodbank,article_organ_A,article_organ_B,article_organ_C}.

Applied as an economic instrument to the context of traffic demand management, Karma mechanisms can be a valuable complement to monetary markets.
Karma-driven economic instruments can overcome equity issues and contribute to public acceptance and support of traffic demand management.
Besides, \textbf{Karma does not create additional, financial costs or taxes to the users}, which overcomes the unwillingness to pay for road usage.
Karma intrinsically embodies a fairness enforcing scheme, that balances consumption and production of resources, and ultimately controls a sustainable usage of public goods.




\section{Materials and Methods} \label{sec:methods}

\subsection{Game Theoretic Formalism}

In previous works, Karma was rather described verbally, which impedes systematic, quantitative analysis.
Modelling Karma as a game is useful, as it allows to predict user behaviour, and to simulate Karma economies as multi-agent-systems~\cite{riehl2024karma}.
In this chapter we outline the game theoretic formalism used for our software framework, and explain how the agent behaviour can be predicted using the social state of the stationary Nash equilibrium for dynamic population games.
A detailed discussion of implicit assumptions of the Karma Game can be found in Appendix~\ref{sec:game-assumptions}.

\subsubsection{Notation}
We define indexes as non-capitalized letters, e.g.\ $i$. 
We denote sets as calligraphic letters, e.g.\ $\mathcal{A}$.
We denote scalars as non-capitalized, indexed letters, e.g.\ $\tau_i$ or $u_i^t$.
We denote functions as capitalized letters, with discrete arguments in square brackets and continuous arguments in round brackets e.g.\ $A[b](c) = d$.
We denote probabilistic functions as Greek letter with subscript p, e.g.\ $\pi_p$.
We denote $Pr(a|b)$ to describe the conditional probability of an event $a$ given $b$.
Table~\ref{tbl-game-notation} summarizes the notation used to describe the Karma game, and to model the resource allocation problem for the software framework (see next chapter).

\begin{table}
    \begin{tabular}{cl}
        \toprule
        \textbf{Symbol} & \textbf{Description} \\
        \midrule
        
        \multicolumn{2}{l}{\textbf{Indexes \& Scalars}} \\
        $i$ & Index of agent in population \\
        $j$ & Index of participant in interaction \\
        $t$ & Index of time/epoch of the game \\
        $e$ & Index of interaction \\
        $n$ & Number of agents in population \\
        
        \multicolumn{2}{l}{\textbf{Sets}} \\
        $\mathcal{N}$ & Set of agents in population \\
        $\mathcal{T}$ & Set of possible agent types \\
        $\mathcal{U}$ & Set of possible urgency levels \\
        $\mathcal{K}$ & Set of possible Karma balances \\
        $\mathcal{J}$ & Set of participants in interaction \\
        $\mathcal{A}_k$ & \parbox[t][][t]{5.5cm}{ Set of possible actions for participant (with Karma balance $k$) } \\
        $\mathcal{O}$ & \parbox[t][][t]{5.5cm}{ Set of possible outcomes from an interaction } \\
        $\mathcal{B}_e$ & \parbox[t][][t]{5.5cm}{ Set of participants' actions (in interaction $e$) } \\
        
        \multicolumn{2}{l}{\textbf{Agent state}} \\
        $\tau_i$ & Type \\
        $u_i^t$ & Urgency level \\
        $k_i^t$ & Karma balance \\
        
        \multicolumn{2}{l}{\textbf{Interaction}} \\
        $a_j^e$ & Action of participant $j$ in encounter $e$ \\
        $o_e$ & Outcome of interaction \\
        $o_j^e$ & Outcome of interaction $e$ for participant $j$ \\
        
        \multicolumn{2}{l}{\textbf{Modelling (Probabilistic Functions)}} \\
        $\Theta_p[o, \mathcal{B}]$ & \parbox[t][][t]{5.5cm}{ Probability of outcome $o$ given the participant actions $\mathcal{B}$ } \\
        $\Omega_p[k_j^{t+1}, k_j^t, \mathcal{B}_e, o_j^e]$ & \parbox[t][][t]{5.5cm}{ Probability of next Karma $k^{t+1}$ given current Karma $k^{t}$, participant's actions $\mathcal{B}_e$  and the participant's outcome $o_j^e$ } \\
        $\Psi_p[\tau, u_j^{t+1}, u_j^t, o_j^e]$ & \parbox[t][][t]{5.5cm}{ Probability of next urgency $u_j^{t+1}$ given current urgency $u_j^{t}$, outcome $o_j^e$, type $\tau$ } \\
        
        \multicolumn{2}{l}{\textbf{Modelling (Logic Functions)}} \\
        $C[u,o]$ & \parbox[t][][t]{5.5cm}{ The immediate costs for a given urgency level and outcome } \\
        $T[\tau]$ & \parbox[t][][t]{5.5cm}{ The discount factor for a given agent type (of temporal preference) } \\
        $Z$ & \parbox[t][][t]{5.5cm}{ Karma overflow account} \\
        $\delta k_i^t$ & Karma payment (positive means receiving) \\

        \multicolumn{2}{l}{\textbf{Social State}} \\
        $\pi_p[\tau,u,k,a]$ & \parbox[t][][t]{5.5cm}{ Probability of action $a$ given the state $\tau,u,k$ } \\
        $d_p[\tau,u,k]$ & \parbox[t][][t]{5.5cm}{ Share of population that has specific type $\tau$, urgency level $u$ and Karma balance $k$ } \\

        \multicolumn{2}{l}{\textbf{Optimization (Intermediate Products)}} \\       
        $\nu_p[a]$ & \parbox[t][][t]{5.5cm}{ Probability of action $a$ (average agent) } \\
        $\gamma_p[o,a]$ & \parbox[t][][t]{5.5cm}{ $Pr(o | a)$ (average agent) } \\
        $\kappa_p[k^{*},k,a]$ & \parbox[t][][t]{5.5cm}{ $Pr(k^{*} | k, a)$ } \\
        $\xi[u,a]$ & \parbox[t][][t]{5.5cm}{ Immediate cost for agent } \\
        $\rho_p[\tau,u^{*},k^{*},u,k,a]$ & \parbox[t][][t]{5.5cm}{ $Pr(u^{*},k^{*} | k, u, a, \tau)$ } \\
        $R[\tau,u,k]$ & \parbox[t][][t]{5.5cm}{ Expected immediate cost } \\
        $P_p[\tau,u^{*},k^{*},u,k]$ & \parbox[t][][t]{5.5cm}{ $Pr(u^{*},k^{*} | k, u, \tau)$ } \\
        $V[\tau,u,k]$ & \parbox[t][][t]{5.5cm}{ Expected infinite horizon cost } \\
        $Q[\tau,u,k,a]$ & \parbox[t][][t]{5.5cm}{ Single-stage deviation reward } \\
        $\widetilde{\pi}_p[\tau,u,k,a]$ & \parbox[t][][t]{5.5cm}{ Perturbed best response policy } \\

        \multicolumn{2}{l}{\textbf{Optimization (Hyper Parameter)}} \\
        $\eta$ & \parbox[t][][t]{5.5cm}{ Change speed of $\pi_p$ relative to $d$ } \\
        $\varpi$ & \parbox[t][][t]{5.5cm}{ Change speed of $\pi_p$ } \\
        $\lambda$ & \parbox[t][][t]{5.5cm}{ Greediness when calculating $Q$ } \\

        \bottomrule
    \end{tabular}
    \caption{
        \textbf{Notation for Karma game and modelling.} 
        }
    \label{tbl-game-notation}
\end{table}

\subsubsection{Karma as a game}

\begin{leftbar}
$Karma$ is as a repeated, stochastic, dynamic population $Game$, represented by a $tuple$ $G$ of parameters.
$G=\left \langle \mathcal{N}, \mathcal{T}, \mathcal{U}, \mathcal{K}, C, T, \Theta_p, \Omega_p, \Psi_p, \pi_p, d \right \rangle$
\end{leftbar}

In a Karma game, there is a population of $n$ agents. For each point in time $t$ (epoch) of the game, each agent $i$ of the population $i \in \mathcal{N} = \{ 1, ..., n\}$ possesses a state consisting of a type $\tau_i$, an urgency level $u_i^t$, and a Karma balance $k_i^t \geq 0, k_i(t) \in \mathcal{K}$.

During each epoch $t$, a subset of agents $\mathcal{J} \subset \mathcal{N}$ is encountering a competition $e$ for a resource at hand (interaction). 
Each participating agent $j \in \mathcal{J}$ during the encounter makes a decision based on its own type $\tau_j$, urgency level $u_j^t$, and Karma balance $k_j^t$ independently from each other (as the state information of each agent is private, invisible to others) using the policy $\pi_p = Pr(a|\tau,u,k)$, and thus executes an action $a_j^e \sim \pi_p$ during the encounter. 
As a result of the interaction there is an outcome $o_j^e$, that determines how the resource is distributed for agent $j$ (e.g.\ if it receives the resource). 
The outcome (resource allocation) affects the future states of the participating agents $u_j^{t+1}, k_j^{t+1} \forall j \in J$. 
The subset $\mathcal{J}$ is randomly and independently chosen from the population.
In one epoch, there can be multiple interactions.

The agent type $\tau_i \in \mathcal{T}$ is time invariant and represents the agent's type of temporal preference.
The agent urgency level $u_i^t \in \mathcal{U}$ determines the immediate costs (rewards) the agent experiences after the interaction of an encounter $e$ by the outcome $o_e$.
The immediate cost an agent experiences is described by function $C[u,o]$, that maps the urgency level and interaction outcome to immediate costs.
The discount factor function $T[\tau]$ maps the discount factor to a temporal preference type.
The discount factor $ T \left [ \tau \right ] \in \left \{ \mathbb{R} \vert 0 \leq T \left [ \tau \right ] < 1 \right \} $ 
indicates how much the agent trades off future costs (rewards) over present costs (rewards).
$T[\tau]=0$ would not consider future costs at all, while $T[\tau]=1$ would not consider present costs at all.

The possible actions $a_j^e$ that the agent can choose from during the encounter $e$, is determined by its Karma balance $a_j^e \in \mathcal{A}_{j,e} = \{1 ... k_j \}$.
The decision making process of the agent to choose an action $a_j^e$ from the set of possible actions $\mathcal{A}_{j,e}$ based on its own, private state $\tau_j$, $u_j$ and $k_j$ is modelled by the policy $\pi_p$. The policy $\pi_p$ is a probabilistic function that maps from a given private state to a probability distribution over actions $\pi_p[\tau_j,u_j,k_j,a_j] = Pr(a_j | \tau_j,u_j,k_j)$. 

The actions of all participants $\mathcal{B}_e = \{a_j^e \forall j \in \mathcal{J} \}$ cause one outcome $o_e$, which is the vector of outcomes for each participant $j$ $o_j^e \in \mathcal{O}$ that is determined by the probabilistic outcome function $\Theta_p$. 
The outcome function $\Theta_p[o_e, \mathcal{B}_e] = Pr(o_e | \mathcal{B}_e)$ maps from all participants' actions $B_e$ to the probability of different possible outcomes.
The outcome $o_e$ of interaction $e$ affects the participant's next Karma balance $k_j^{t+1}$ according to a probabilistic function $\Omega_p$, that represents the Karma transition probability, referred to as the Karma payment rule: $\Omega_p [k_j^{t+1}, k_j^t, \mathcal{B}_e, o_j^e] = Pr( k_j^{t+1} | k_j^t, \mathcal{B}_e, o_j^e)$.
The outcome $o_e$ of interaction $e$ affects the agent's next urgency level $u_j^{t+1}$ according to a probabilistic transition function $\Psi_p$, that represents the urgency transition probability, respectively the $\tau$-dependent urgency process: $\Psi_p [\tau, u_j^{t+1}, u_j^t, o_j^e] = Pr( u^{t+1} | u^t, o_j^e, \tau)$.

The distribution of types, urgency levels, and Karma balances in the population is the state distribution $d_p$. 
$d_p[\tau,u,k]$ describes the share of the population that has a specific type $\tau$, urgency level $u$, and Karma balance $k$. 
Together with the policy, the state distribution is called the social state of the Karma game $(\pi_p, d_p)$.

\subsubsection{Agent behaviour prediction}

\begin{figure}[H]
    \centering
    \includegraphics[width=\linewidth]{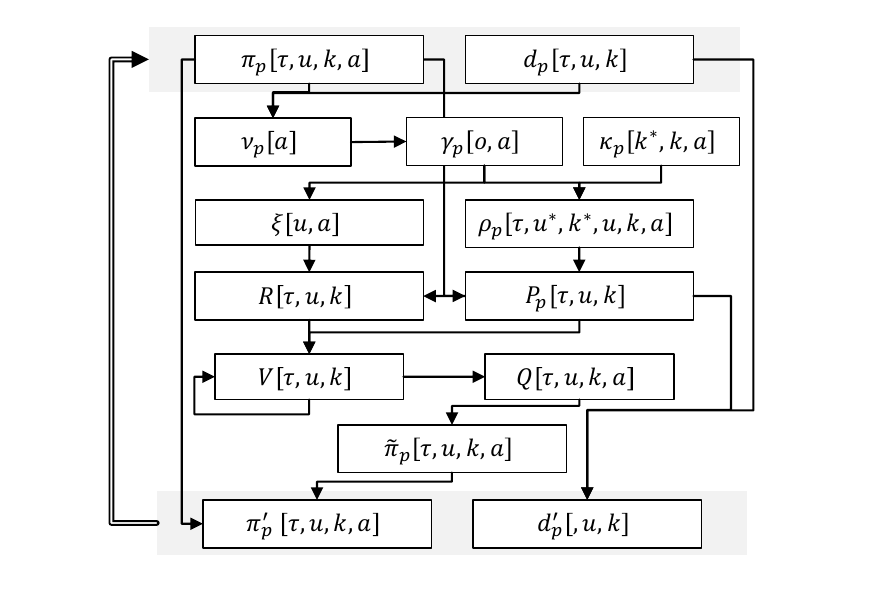}
    \caption{
        \textbf{Stationary Nash Equilibrium Calculation.} \\  {\scriptsize }
        The consumption behaviour of rational, selfish individuals in Karma economies can be predicted by calculating the stationary Nash Equilibrium. The stationary Nash Equilibrium~\cite{article_600} consists of two components: (i) a probabilistic policy matrix $\pi_p$, that describes how much an individual would be willing to pay $a$, given his temporal preference type $\tau$, urgency $u$, and Karma balance $k$; (ii) a population distribution $d_p$, that describes the share of the population with given $\tau$, $u$, and $k$.        The stationary Nash Equilibrium can be calculated in an iterative process via intermediate results, as shown in this figure. 
    }
    \label{fig-optim-calc}
\end{figure}

In order to simulate the Karma mechanism as a multi-agent-system, it is of crucial importance to predict the behaviour $\pi_p[\tau,u,k,a]$ of agents.
Usually one assumes that each agent is rational and acts in its best self-interest, meaning an agent achieves the best possible outcome for itself.
The \textbf{rational behaviour of an agent is described by the optimal policy} $\pi_p^{*}[\tau,u,k,a]$.
The \textbf{Nash-equilibrium of a game describes the optimal policy}, where no agent can improve its situation from deviating from the policy~\cite{article_567,article_600,article_1001}.
A population of rational agents following this optimal policy will lead to a stationary state distribution $d^{*}_{p}$.
For dynamic population games, the concept of a stationary Nash-equilibrium describes the social state, that consists of the optimal policy and stationary state distribution $(\pi_{p}^{*}, d_{p}^{*})$.
At least one stationary Nash-equilibrium for each Karma game is guaranteed to exist when the circulating amount of Karma is preserved~\cite{article_567,article_600,article_1001}.
The optimal social state can be calculated in an iterative way, as outlined in Fig.~\ref{fig-optim-calc}, by computing intermediate products.

$\nu_p[a]$ represents the probability distribution of an average agent's actions.

\begin{equation}
\nu_p[a] = \sum_{\tau,u,k} d_p[\tau,u,k] \pi_p[\tau,u,k,a]
\end{equation}

$\gamma_p[o,a]$ represents the probability of an interaction outcome for an agent given its action $a$, and the action(s) of randomly chosen opponent(s) ${a}'$.

\begin{equation}
\gamma_p[o,a] = \sum_{{a}'} \nu_p[{a}'] \Theta_p[o,a,{a}']
\end{equation}

$\kappa_p[k^{*},k,a]$ represents the probability, that an agent will have a Karma balance $k^{*}$ after the interaction, given a previous Karma balance $k$ and action $a$.
Depending on the $\Omega_p[k_j^{t+1}, k_j^t, \mathcal{B}_e, o_j^e]$ logic defined in the Karma game, e.g.\ highest vs. second highest bid wins, winner pays bid to peer vs. to society, Karma redistribution etc., the modelling of this function is a non-trivial task (see Appendix~\ref{sec:game-assumptions}). 

$\xi[u,a]$ represents the immediate costs an agent of type $\tau$, with urgency level $u$, Karma balance $k$ performs action $a$ in an interaction.

\begin{equation}
\xi[u,a] = -  \sum_{o} C[u,o] \gamma_p[o,a]
\end{equation}

$\rho_p[\tau,u^{*},k^{*},u,k,a]$ represents the probability, that an agent of type $\tau$ will have an urgency level $u^{*}$ and Karma balance $k^{*}$ after the interaction, given a previous urgency level $u$, Karma balance $k$, and action $a$.

\begin{equation}
\rho_p[\tau,u^{*},k^{*},u,k,a] = \sum_{o} \Psi_p[\tau, u^{*}, u, o] \gamma_p[o,a] \kappa_p[k^{*},k,a]
\end{equation}

$R[\tau,u,k]$ represents the expected immediate cost for an agent of type $\tau$, urgency level $u$, and Karma balance $k$ that follows the policy $\pi_p[\tau,u,k,a]$.

\begin{equation}
R[\tau,u,k] = \sum_{a} \pi_p[\tau,u,k,a] \xi[u,a]
\end{equation}

$P_p[\tau,u^{*},k^{*},u,k]$ represents the probability, that an agent of type $\tau$ will have an urgency level $u^{*}$ and Karma balance $k^{*}$ after the interaction, given a previous urgency level $u$ and Karma balance $k$, assuming that the agent follows the policy $\pi_p[\tau,u,k,a]$.

\begin{equation}
P_p[\tau,u^{*},k^{*},u,k] = \sum_{a} \pi_p[\tau,u,k,a] \rho_p[\tau,u^{*},k^{*},u,k,a]
\end{equation}

$V[\tau,u,k]$ represents the expected infinite horizon cost for an agent of type $\tau$, urgency level $u$ and Karma balance $k$.
$V[\tau,u,k]$ can be computed by the recursive, Bellman-equation as shown below, and is guaranteed to converge due to contraction-mapping. 

\begin{equation}
V[\tau,u,k] \leftarrow R[\tau,u,k] + T[\tau] \sum_{u^{*},k^{*}} P_p[\tau, u^{*},k^{*},u,k] V[\tau,u^{*},k^{*}]
\end{equation}

$Q[\tau,u,k,a]$ is the single-stage deviation reward, when deviating from the current iteration's policy. This intermediate computation product highlights where deviating from the current iteration's policy is profitable and guides the process of improving the policy. In the stationary Nash equilibrium, $Q$ would not deviate from the policy any more.

\begin{equation}
Q[\tau,u,k,a] = \xi[u,a] + T[\tau] \sum_{u^{*},k^{*}} \rho_p[\tau,u^{*},k^{*},u,k,a]V[\tau,u,k]
\end{equation}

$\widetilde{\pi}_p[\tau,u,k,a]$ represents the perturbed best response policy. The hyper parameter $\lambda$ controls for how strong (greedy) the single-stage deviation reward of this iteration should be taken into account when improving the policy.

\begin{equation}
\widetilde{\pi}_p[\tau,u,k,a] = \frac{\exp(\lambda Q[\tau,u,k,a] - \max_{{a}'} \lambda Q[\tau,u,k,{a}'])}{\sum_{{a}'} \exp(\lambda Q[\tau,u,k,a] -  \max_{a^{*}} \lambda  Q[\tau,u,k,a^{*}])}
\end{equation}

Finally, the update of social state consists of two steps based on the computed, intermediate products $P$ and $\widetilde{\pi}_p$.
Two hyper parameters, control the speed of change for the distribution ($\varpi$), and for the policy relative to the distribution ($\eta$ ).

\begin{equation}
\pi_p \leftarrow (1 - \eta \varpi) \pi_p + (\eta \varpi) \widetilde{\pi}_p
\end{equation}

\begin{equation}
d_p \leftarrow (1 - \varpi ) d_p + (\varpi ) d_p P_p
\end{equation}

Further information on the computation of the Stationary Nash Equilibrium can be found in Appendix~\ref{sec:optimizers}.

\subsection{Software Framework}

In this work, we present a well-documented, open-source, software framework (Python, PEP8, GPL 3.0) to equip the reader with the toolset to apply Karma as a resource allocation mechanism.
It implements the game theoretic formalism, allows for the computation of the stationary Nash-equilibrium, provides a rich template library to model Karma, and enables the simulation of Karma as a multi-agent-system. 

To use the software, users will need to follow a three-step approach: (i) defining their Karma game (modelling), (ii) predicting the behaviour of market participants (optimization), and (iii) simulation of a multi-agent-system Karma economy (simulation).
In the following we will outline how the reader can model a resource allocation problem, how the optimal policy at the stationary Nash equilibrium can be computed using the optimization module, and finally how to simulate the resource allocation as a multi-agent-system.
For more information and details, please refer the to GitHub page and documentation.

\subsubsection{Modelling}

In order to model a Karma resource allocation problem with the software framework, the user needs to specify general parameters, logic functions (for simulation) and probabilistic functions (for optimization). 
The probabilistic functions need to be provided to capture how an individual agent would model reality and make decisions, in order to predict the optimal behaviour of a rational agent, by computing the stationary Nash equilibrium. 
The logic functions need to be defined in order to simulate the Karma game as a multi-agent-system.
The software framework offers a rich library of predefined templates and examples for the probabilistic and logic functions.

To begin with, the user needs to specify the average initial Karma, number of agents $n$, number of participants in an interaction $\| \mathcal{J} \|$, initial distribution $d_p[\tau,u,k]$, set of temporal preferences $\mathcal{T}$, set of urgency levels $\mathcal{U}$, set of valid Karma balances $\mathcal{K}$, and set of possible interaction outcomes $\mathcal{O}$.

Next, the user needs to specify probabilistic functions, namely 
the probabilistic outcome function $\Theta_p$ ($Pr(o | \mathcal{B})$), 
the probabilistic Karma transition function $\Omega_p$ ($Pr(k^{t+1} | k^{t}, \mathcal{B}_e, o_{e,j})$), and 
the probabilistic urgency transition function $\Psi_p$ ($Pr(u^{t+1} | u^{t},o, \tau)$).

Afterwards, the user needs to specify logic functions, that determine payments between participants $\delta k_i$ and/or a Karma overflow account $Z$ that can be redistributed: 
the cost function $C([u_j,o_j] \rightarrow \mathbb{R})$, 
the temporal preference function $T$ ($[\tau_i] \rightarrow [0;1] \in \mathbb{R}$), 
the outcome function ($\mathcal{B}_e \rightarrow o_e$),
the payment function ($[a_j, o_j] \rightarrow [\left \{ \delta k_i \forall i \in \mathbb{N} \right \}, Z]$), 
the urgency transition function ($[u_i,o_i] \rightarrow u_i$), 
the overflow distribution function ($\left [\{ k_i \forall i \in \mathbb{N} \right \}, Z] \rightarrow \left \{ \delta k_i \forall i \in \mathbb{N} \right \}$), and 
the Karma redistribution function ($\left \{ k_i \forall i \in \mathbb{N} \right \} \rightarrow \left \{ \delta k_i \forall i \in \mathbb{N} \right \}$).

\begin{table}[H]
    \centering
    \begin{tabular}{|l|l|}
        \toprule
        \textbf{Design Parameter} & \textbf{Modelling Aspect} \\
        \midrule
        
        & \\
        \textbf{Currency} & \\
        \hspace*{0.5cm} Parity & $\mathcal{A}_k$, $\mathcal{O}$ \\
        \hspace*{0.5cm} Balance limits & $\mathcal{K}$ \\
        \hspace*{0.5cm} Amount control & Constant \\
        \hspace*{0.5cm} Initialization & Initial distribution \\
		\hspace*{0.5cm} Redistribution & Redistribution function \\
			        
        & \\
        \textbf{Interaction} & \\
        \hspace*{0.5cm} Price control & Payment function \\
        \hspace*{0.5cm} Price limits & $\mathcal{A}_k$, Payment function \\
        \hspace*{0.5cm} Resource provision & Problem-specific \\
        \hspace*{0.5cm} Resource allocation & Outcome function \\
        \hspace*{0.5cm} Counter-party & $\mathcal{J}$ \\
        \hspace*{0.5cm} Peer selection & (Options by simulation module) \\
        \hspace*{0.5cm} Decision making & Optimal policy $\pi_p[\tau,u,k,a]$ \\
        \hspace*{0.5cm} Urgency process & $\Psi_p$, Urgency transition function \\
        \hspace*{0.5cm} Temporal preference & $T[\tau]$, Discount factors \\

        & \\
        \textbf{Transaction} & \\
        \hspace*{0.5cm} Payment amount & Payment function \\
        \hspace*{0.5cm} Payment receiver & Payment function, \\
        & Overflow Distribution function \\
        \hspace*{0.5cm} Karma gain \& & Payment function, \\
        \hspace*{0.5cm} Karma loose & Overflow distribution function, \\
        &  Redistribution function \\
        
        & \\
        \hline
    \end{tabular}
    \caption{\textbf{Karma Design and Modelling}}
    \label{table_design_modell}
\end{table}

Table~\ref{table_design_modell} connects the design parameters of the Karma mechanism with the modelling aspects of the software framework.
Please note, the framework encodes the outcome $o_{e,j}=0$ as not receiving a resource (costs appear) and $o_{e,j}=1$ as receiving the resource (no costs appear).

\subsubsection{Optimization}

In this section, we showcase the optimization process, which represents the computation of the Karma Game's stationary Nash-equilibrium, as described in Algorithm~\ref{alg:alg1}. For further information on the optimization process and alternatives, review Appendix~\ref{sec:optimizers}.


\begin{algorithm}
\caption{Behaviour prediction}\label{alg:alg1}
\begin{algorithmic}[1]
\State Init social state
\While{not $AbortionCriterion$}
    \State Adjust state space
    \State Validate social state
    \State Compute intermediate products (in this order)
    \State Update social state
\EndWhile
\end{algorithmic}
\end{algorithm}


\textbf{Init social state: }
To initialize the state distribution, one needs to define an initial distribution across agent types, urgency levels and Karma balances.
The distribution will determine the average Karma amount in the population, which needs to stay constant over the runtime of the algorithm.
To initialize the policy, the software framework offers three possible initializations by default: ``bottom'', ``even'' and ``top'', that represents initial policies in which the agents always bid 0 (bottom), always bid the maximum Karma amount (top) or an even distribution between them (even). 
We recommend ``even'' as we observed the convergence to proceed faster.

\textbf{AbortionCriterion: }
As mentioned in the game theoretic model, we employ an iterative approach to calculate the stationary Nash equilibrium.
Over many iterations, the difference of the social state after and before the iteration decreases.
While the algorithm will converge towards the stationary Nash equilibrium, at some point, when the precision of the social state is sufficient, one can abort the computation.
A certain abortion criterion defines the sufficiency in this context.
In our software framework, we recommend the user to have a dual abortion criterion: maximum number of iterations, and a convergence threshold for the differences of social states between iterations.

\textbf{Adjust state space: }
While the Karma game, as defined in this work, could have an infinite state (possible Karma balances $\mathcal{K}$) and thus action space (possible actions $\mathcal{A}$), in practice it is impossible and unnecessary to calculate $d_p$ and $\pi_p$ for infinite spaces. 
In the software framework, we store $d_p$ and $\pi_p$ in arrays (tensors) of finite dimensions.
We define initial state and action space based on the average initial Karma, and then dynamically expand the spaces when certain conditions are met, which make expansion necessary.

The initial state and action space need to be set by the user. 
We recommend an initial action space with a size equal to the average initial Karma, and
an initial state space with a size equal to four times the average initial Karma (as within the first iterations the Karma distribution at the average initial Karma sinks and spills over to neighbouring Karma balances on the left and rights).

The action space is expanded, if the sum of the policy's action probabilities of the boundary action (highest action in the space) across all types, urgencies and Karma balances, exceed a threshold value.
The state space is expanded, if the sum of the distribution's shares for the highest four boundary states (highest Karma balances in the space) across all types, and urgencies, exceeds a threshold value. 
We do so, to make sure there is always enough space for the distribution to expand. 
Based on our computations, we have found the distribution to react sensitive and convergence decelerates significantly if the distribution hits the boundaries.

\textbf{Validate social state: }
The numerical computations of the algorithm with decimal floating point numbers can cause rounding errors that accumulate over the iterations.
Thus, it is important to regularly validate whether the social state is correct, and if not, to correct through normalization.
The $d[\tau,u,k]$ is valid, if: (i) all the shares for different types, urgency levels, and Karma balances add up to 1.0, and (ii) the average Karma balance equals the average initial Karma balance:

\begin{equation}
\sum_{\tau,u,k} d_p[\tau,u,k] = 1.0 \wedge \sum_{\tau,u,k} k \times d_p[\tau,u,k] = \rm{const.}
\end{equation}

The policy $\pi_p[\tau,u,k,a]$ is valid, if the probabilities of all actions for a given agent type, urgency level, and Karma balance adds up to 1.0:

\begin{equation}
\sum_{a} \pi_p[\tau,u,k,a] = 1.0 \;\; \forall \;\; \tau, u, k
\end{equation}


\textbf{Compute intermediate products: }
The intermediate products are calculated using an evolutionary, best-response dynamic, as described in the game theoretic formalization.
Please note, the iterative calculation of $V$ is initialized as $V[\tau,u,k]=0$, and aborted based on two criteria: (i) maximum number of iterations, (ii) convergence threshold.

\textbf{Update social state: }
The update of the social state follows the elaboration in the game theoretic formalization.
The hyper parameters can be tweaked for specific optimization problems, and also changed adaptively over the iterations to accelerate convergence.
Based on our experiences, we recommend the values $\lambda=1000, \varpi=0.20, \eta=0.50$ as a good set of hyper parameters to start with.

\subsubsection{Simulation}
The time-discrete simulation of Karma as a multi-agent system is outlined in Algorithm~\ref{alg:alg2}.
The implemented simulator can be integrated seemlessly with other simulators, and offers storage and computation for all Karma related population values.
At all times, the software framework records the population related information (type, urgency level, Karma balance, cumulated costs, number of encounters), and provides useful methods to retrieve information on the simulation progress.

\begin{algorithm}
\caption{Simulation}\label{alg:alg2}
\begin{algorithmic}[1]
\State Init social state
\While{not $AbortionCriterion$}
    \State Begin epoch
    \State Execute interactions
    \begin{algsubstates}
        \State Participant selection
        \State Decision making process
        \State Determine outcome
        \State Karma transactions (payments \& overflow)
    \end{algsubstates}
    \State Close epoch
    \begin{algsubstates}
        \State Urgency transition
        \State Karma overflow distribution
        \State Karma redistribution
    \end{algsubstates}
\EndWhile
\end{algorithmic}
\end{algorithm}


\textbf{Init social state: }
In addition to the game parameters specified during modelling and optimization, the number of agents $n$, an initial state distribution $d[\tau,u,k]$, and the optimal policy $\pi_p[\tau,u,k,a]$ need to be provided. 
By default, an initial state distribution will be derived from the computed stationary Nash equilibrium, in order to simulate a multi-agent-system that already is in its steady state. 
However, there are options to initiate the system by equally distributing Karma units.

\textbf{AbortionCriterion: }
The simulation happens in discrete periods of time (epoch), and can be repeated until a user-defined abortion criterion is met. Each epoch consists of three computation steps.

\textbf{Begin epoch: }
The first step of an epoch is to record of the states before interactions.

\textbf{Execute interactions: }
The second step of an epoch consists of executing one or multiple interactions. 
Each interaction requires a list of participating agents, to predict the actions of the participants based on the optimal policy, and to compute an outcome of the interaction based on the participant actions.
The outcome of the interaction will then cause the Karma transaction which includes the update of costs, and Karma balances.
The list of participating agents could originate from a domain specific simulator, or could it be generated randomly.

\textbf{Close epoch: }
The third step of an epoch includes the update of states: urgency transition, Karma overflow distribution and Karma redistribution (and internal recording).

\begin{figure} [ht!]
    \centering
    \includegraphics[width=\linewidth]{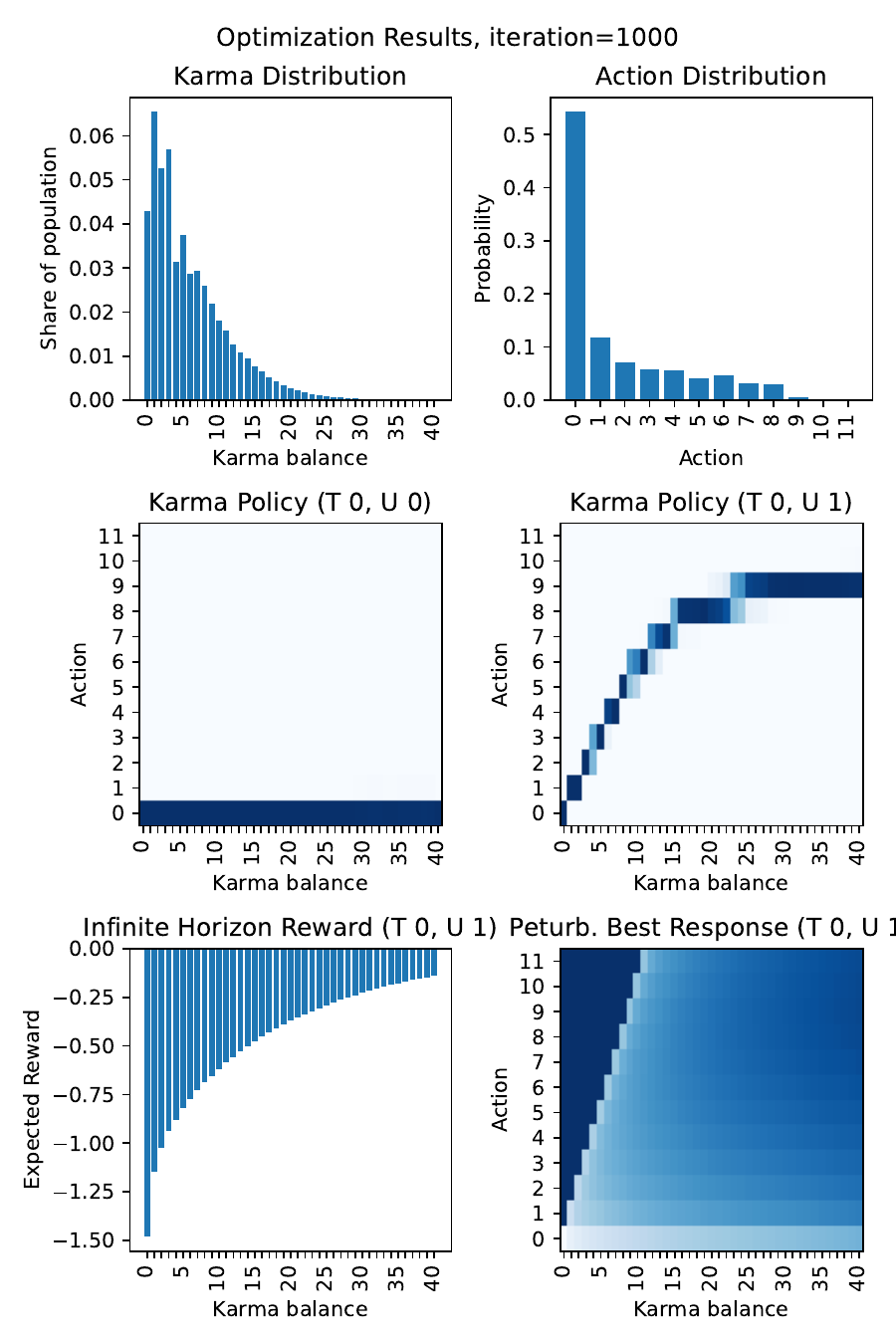}
    \caption{
        \textbf{Social State in Stationary Nash Equilibrium.} \\ {\scriptsize }
        The results of calculating the stationary Nash Equilibrium for an exemplary Karma auction, after thousand iterations. We can observe that the Karma balance across the population follows a t-like distribution (top left), and that the action of a random encountering competitor will be a zero bid with a chance of more than 50\% (top right). The bids (action) for individuals, that do not experience any costs (u=0), are always zero bids. Rather, they save their Karma for an urgent situation (middle left). The bids for individuals, that experience costs when not winning the auction (u=1), are higher for larger Karma balances, but not higher than 9, as the chance of meeting a random competitor with a larger bid are almost zero (middle right). The infinite horizon reward (bottom left) shows, that the expected costs (negative rewards) for an individual depend on its Karma balance. The perturbed best response (bottom right) exhibits, that the policy above (middle right) is the best possible bidding strategy, and that one cannot improve its own outcome by deviating from it.
        }
    \label{fig-optim-results}
\end{figure}

\subsection{Computation Example}

\subsubsection{Modelling}
In this example, we will discuss the optimization and simulation of a resource allocation problem, where there is one type of user with a temporal preference of 0.8, and two urgency levels with costs 0 and 3. 
The initial Karma balance is set to 6. 
In terms of logic, the highest bidding participant in an interaction of two agents wins, and pays the bid to the peer. 
Also, let us assume there is no Karma redistribution.

\subsubsection{Optimization}
Figure~\ref{fig-optim-results} shows the results of the optimization process after 1000 iterations.
Starting from an average initial Karma of 6, we can observe, that the state space expanded up to a Karma balance of 40 units, and the action space expanded to 11 units for bidding (more is irrational for the given set-up) (Figure~\ref{fig-optim-results}, first row).
The distribution $d_p[\tau,u,k]$ of Karma across the population is right skewed, meaning that there is no incentive to hoard large amounts of Karma.
The Karma policies (Figure~\ref{fig-optim-results}, second row) show, that if there is an urgency of zero $\pi_p[\tau,u=0,k,a]$, and thus no costs incurred with not getting the resource, there is little incentive to bid anything except for 0, in order to save the Karma units for situations when the urgency is not zero.
If there are costs incurred to not getting the resource however $\pi_p[\tau,u=1,k,a]$ we can observe a bidding behaviour, up to 9 Karma units. Above that, it doesn't make sense for the participant to bid higher and rather save its Karma units, as there are almost no agents owning more than 20 Karma units, and no competitor will bid more. Moreover, not all available Karma units are used in bidding in order to save the units for future encounters. These two policies lead to the average action distribution of a randomly selected competitor.
The infinite horizon reward $V[\tau,u,k]$ (Figure~\ref{fig-optim-results}, third row) shows, that the expected reward (negative costs) increases (costs decrease) the larger the Karma balance. This means, that participants with higher Karma balances are more likely to win the auctions.
The single stage deviation $Q[\tau,u,k,a]$ matches almost completely with the derived Karma policy after the 1000 iterations, which means the optimal policy is reached, as there is no benefit anymore in deviating from it.

\begin{figure}
   \centering
    \includegraphics[width=\linewidth]{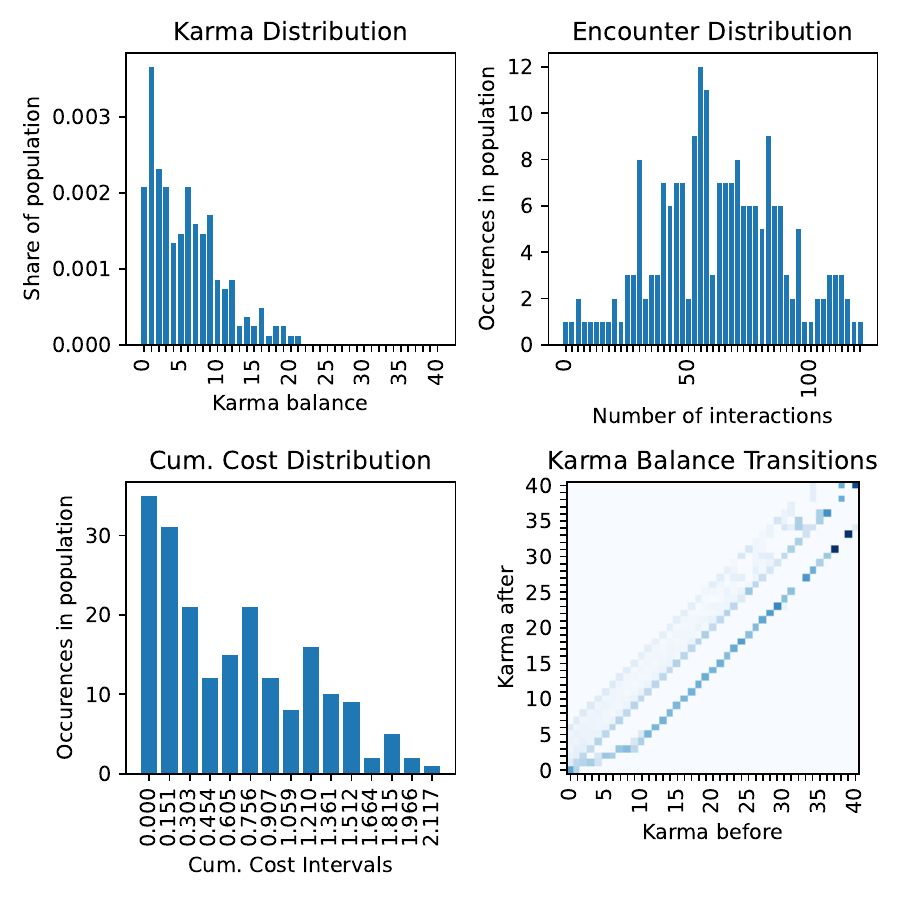}
    \caption{
        \textbf{Simulation as multi-agent-system.} \\ {\scriptsize }
        The simulation of a Karma economy as a multi-agent-system of 200 rational individuals over 10,000 epochs, leads to a Karma distribution similar to the one predicted by the stationary Nash Equilibrium. The encounters (auctions) between two random participants per epoch, happen normally distributed, where on average each individual participated around 75 interactions. The cumulative cost distribution demonstrates, that the chance of experiencing high costs over time is very low. The Karma balance transitions exhibit, that Karma balances either increase by the bid of the opponent, shrink by one's own bid, or stay the same (both zero bids).
        }
    \label{fig-simulation}
\end{figure}

\subsubsection{Simulation}
Figure~\ref{fig-simulation} shows the results of the simulation process after 10,000 epochs with one random interaction of two participants per epoch, and a population of 200 individuals. On average, each agent has had around 75 interactions as a result.
The shape of the Karma distribution resembles the predicted stationary Nash equilibrium. 
Furthermore, we can see that the cumulative costs of the agents are right-skewed, with an average of 0.7, meaning that agents rarely experience severely high costs after multiple interactions. 
The Karma balance transition shows the probabilities that an agent with a Karma balance before (abscissa) ends up with a Karma balance after (ordinate), averaged across all types and urgencies. One can observe three main diagonals, as participants either do not participate in an interaction (remain with the same amount of Karma), or participate and either win the resource (loose Karma) or loose the resource (win Karma).

\section{Case Study} \label{sec:casestudy}

New York City is the most populous and most densely populated city in the United States of America, with an estimated population of 8.3 million people, and a land area of 1.2 square kilometers. 
New York is located at the southern tip of New York State, and divided into five boroughs, that are separated by rivers and the sea.
Culturally and economically, it is one of the most vibrant cities, being home to financial institutions (Wall Street, the New York Stock Exchange), and to headquarters of international corporations and organizations alike (e.g.\ United Nations).
Due to its flourishing economy, New York and especially the borough Manhattan attract many visitors and commuters.

\begin{figure} [ht!]
    \centering
    \includegraphics[width=\linewidth]{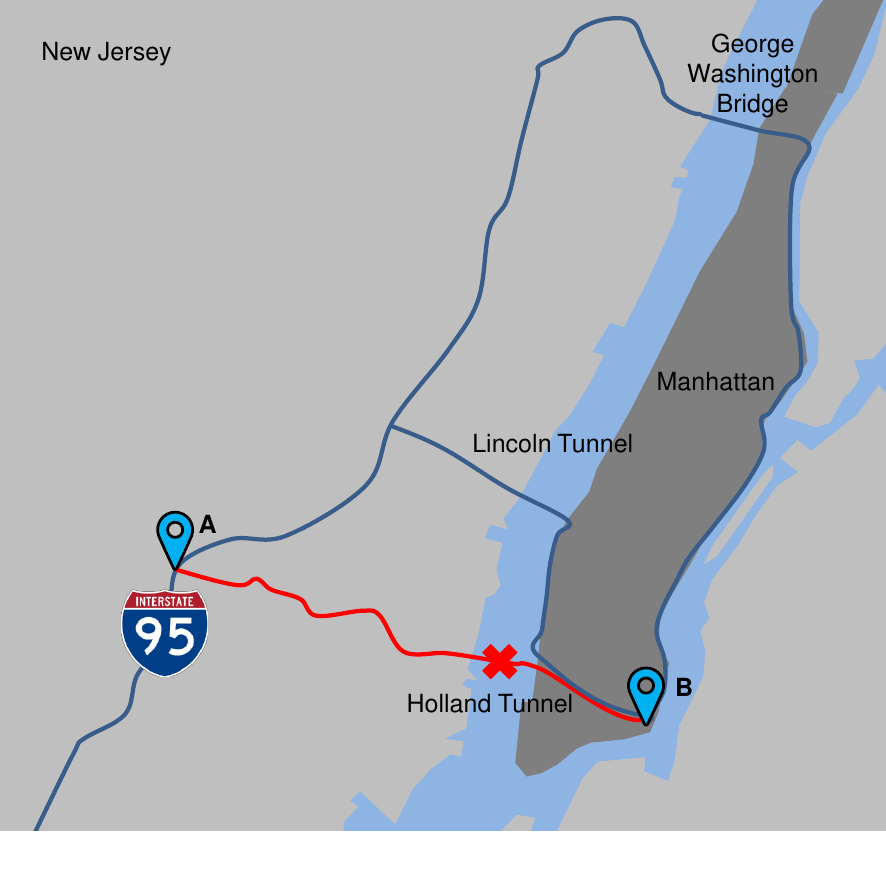}
    \caption{
        \textbf{Case Study: New Jersey and Manhattan (New York)} \\ {\scriptsize }
        Manhattan (New York) attracts a large workforce from the neighbouring state of New Jersey, that travel via interstate 95, and can choose between Holland tunnel, Lincoln tunnel, and George Washington Bridge, to cross the Hudson river. In this case study, we explore the distributional effects of pricing the Lincoln tunnel, assuming the Holland tunnel is closed, and the traffic must distribute across the remaining two routes.
        }
    \label{fig-casestudymap}
\end{figure}


The neighbouring state of New Jersey is home to a large share of Manhattan's commuting workforce, per day around 1.23 million people.
New Jersey and Manhattan are separated by the Hudson river. 
There are mainly three connections, drivers can use to cross the river: the Holland Tunnel (Interstate 78), the Lincoln Tunnel (Route 495), and the George Washington Bridge (Interstate 95), as shown in Fig.\ref{fig-casestudymap}.
Together, these three connections transport more than 493,000 vehicles per day.
The Holland Tunnel consists of two tubes, has an operating speed of 56 km/h, a length of around 2.5 km, 9 lanes, and transports around 89,792 vehicles per day.
The Lincoln Tunnel consists of three tubes, has an operating speed of 56 km/h, a length of around 2.4 km, 6 lanes, and transports around 112,995 vehicles per day.
The George Washington Bridge consists of two decks (levels), has an operating speed of 72 km/h, a length of around 1.4 km, 14 lanes, and transports around 289,827 vehicles per day, making it the world's busiest vehicular bridge.
The connections are separated by 4,37 km and 10.87 km respectively~\cite{newyorkData2016}.

Currently, all bridges and tunnels in New York city follow a unified toll rate scheme by the New York Port authority, with prices per vehicle types, and time (on- and off-peak hours). 
The tolls are collected when entering New York, and not when entering New Jersey.
Special discounts apply for taxis, or ride-sharing vehicles.
Normal passenger vehicles pay between \$13.38 and \$15.38\footnote{New York Port Authority, Toll rates for all Port Authority bridges \& tunnels. 2024 Toll Rates. https://www.panynj.gov/bridges-tunnels/en/tolls.html.}.

Unfortunately, New York is not only known as an attractive city, but also is it known as the city with the worst traffic in America.
On a daily average, vehicles spent 154 seconds per kilometer travelled in New York City (average speed 20km/h) during rush hour, which sums up to 112 wasted hours per year and vehicle due to congestion\footnote{TomTom Traffic Index Ranking 2023. https://www.tomtom.com/traffic-index/ranking/?country=US.}.
Besides, constructions, planned maintenance, scheduled overnight closures, and security incidents cause the regular closure of these important bottlenecks. 

In this case study, let us discuss static road pricing with monetary markets and Karma schemes for a scenario, where only two of the three passages are available.
Let us assume, that there is a fire hazard due to an accident in the Holland Tunnel, and that the tunnel is blocked for a week.
A drive from A to B in our case study map (Fig.~\ref{fig-casestudymap}) would be around 26.55 kilometers (35 minutes free flow) via the Lincoln Tunnel, and 49.89 kilometers (40 minutes free flow) via the George Washington Bridge.
Before closure, it was only around 17.38 kilometers (24 minutes free flow) via the Holland Tunnel.
The traffic that came from Interstate 95 and used to go over the Holland Tunnel, will try to go over the next closest passage nearby: the Lincoln Tunnel.
As s a consequence, the Lincoln Tunnel is facing congestion. 
Therefore, the New York City Port authority decides to price the Lincoln Tunnel higher, and to stop charging for the George Washington Bridge.
Doing so, the authority aims to distribute the additional traffic more efficiently between the two connections.
In order to mitigate congestion, the authority will choose a price that minimizes the total travel time.
We will compare the effects of road pricing between monetary markets and Karma markets.

\begin{figure} [ht!]
    \centering
    \includegraphics[width=\linewidth]{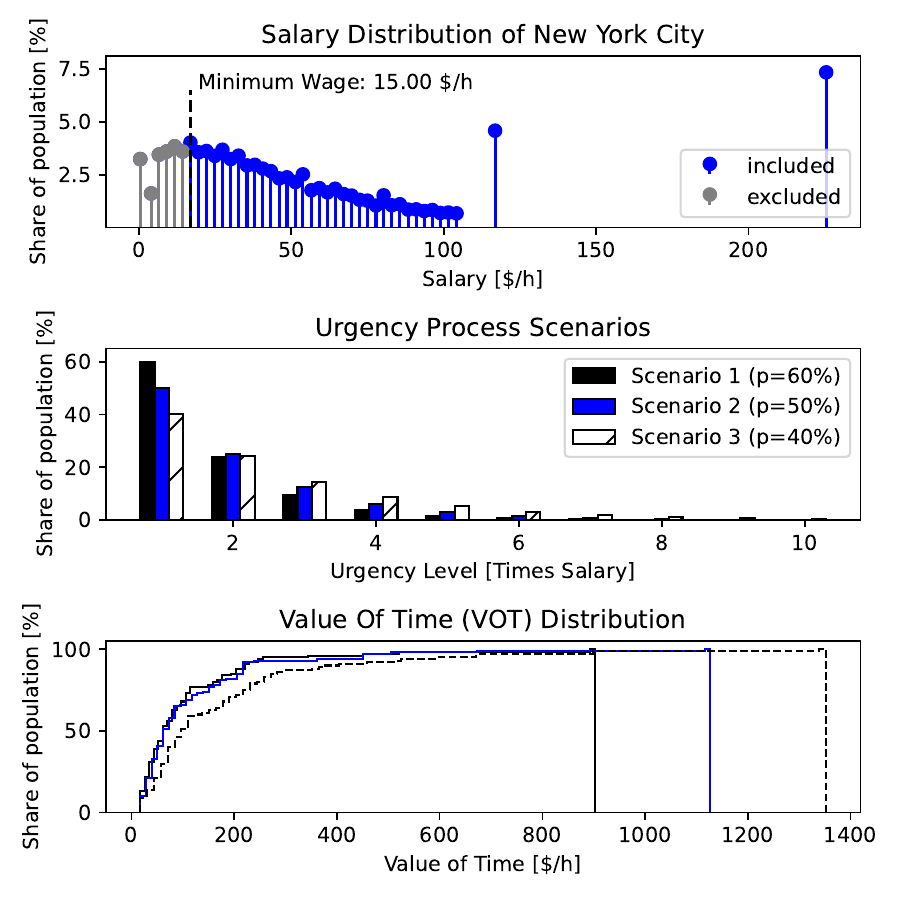}
    \caption{
        \textbf{Case Study: Population Urgency Model} \\ {\scriptsize }
        Combining salary and urgency distribution results in a value of time (VOT) distribution, that can be used to analyse the distributional effects using monetary road pricing. The urgency of an individual represents its willingness to pay n-times its salary for using the Lincoln tunnel. We assume three, geometrically-distributed urgency scenarios for this investigation.
    }
    \label{fig-urgencymodel}
\end{figure}

Fig.~\ref{fig-urgencymodel} depicts the population urgency model.
We model the population with ten urgency levels (1 to 10), where the urgency levels are assumed to be randomly-geometrically distributed in three different scenarios (p=0.6, p=0.5, p=0.4).
The $n$-th urgency level represents delay costs of $n$ times the hourly wage\footnote{The salary distribution of New York city according to 2022 US Census data is assumed: https://en.wikipedia.org/wiki/Household\_income\_in\_the\_United\_States\#Distribution \_of\_household\_income .} (value of time, VOT).

\begin{figure} [ht!]
    \centering
    \includegraphics[width=\linewidth]{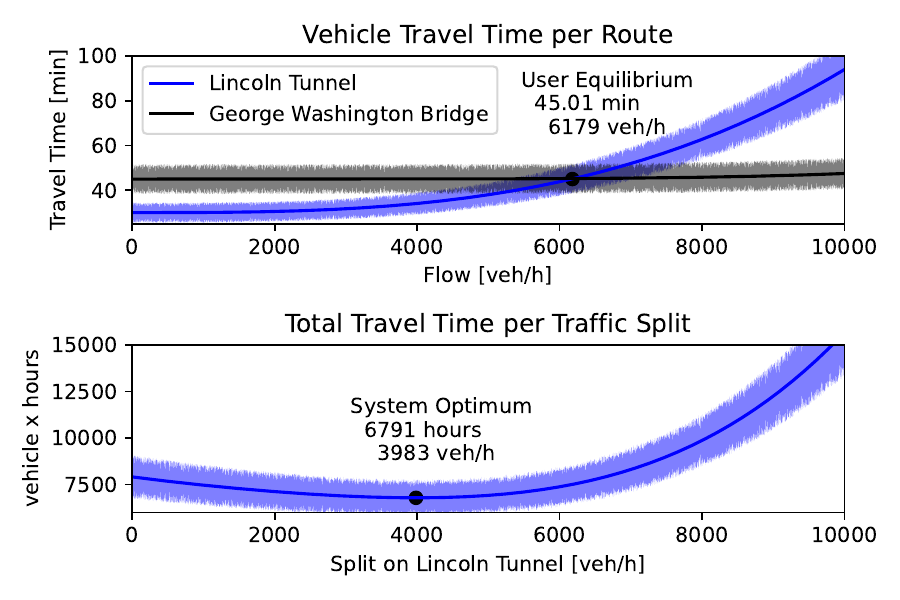}
    \caption{
        \textbf{Case Study: Travel Time Model} \\ {\scriptsize }
        The travel times per route depend on the traffic (flow) on each route. While the Lincoln tunnel is a faster route in general, it reacts more sensitive to higher traffic, and gets much slower due to congestion. The George Washington Bridge is a slower route, but has a higher capacity and therefore does not react to higher traffic flows. From a system perspective, the optimal traffic flow split can be achieved, when the total travel time (vehicle hours) are minimized, meaning around 3983 veh/h on the tunnel route, and the rest via the bridge. However, rational individuals, that try to achieve the best outcome for themselves, lead to a tunnel usage (Wardrop equilibrium), where the travel times of both routes are equal. As a result, the selfish behaviour of rational individuals causes 4.26 minutes of additional travel time per vehicle on average.
    }
    \label{fig-delaymodel}
\end{figure}

Fig.~\ref{fig-delaymodel} depicts the travel time model.
We assume a traffic of 10000 veh/h, which is split across the two routes. 
While the Lincoln Tunnel has a shorter travel time upfront, it gets congested quickly, and after 6000 veh/h it is much slower when compared with the alternative route. 
The route via George Washington Bridge offers slower travel times, but higher capacity and less congestion and delays for even higher flows.
From a system optimal point of view, a minimum total travel time of 6791 vehicle hours (40.75 minutes average travel time) can be achieved, if the total flow (10000 veh/h) splits to 3983 veh/h on the Lincoln route and 6017 veh/h on the George Washington route.
Unfortunately, rational (selfish) individuals would optimize their individual outcome, leading to an user equilibrium (Wardrop equilibrium) at a split of 6169 veh/h on the Lincoln tunnel, as there is no way to improve one's individual outcome by changing anymore.
The split at the user equilibrium (45.01 minutes average travel time) causes 4.26 minutes of additional travel time to every vehicle on average.
With the right pricing of the Lincoln Tunnel, the total travel time could be reduced by almost 10\%.

\section{Results} \label{sec:results}

In this case study we aim to compare the distributional effects of money and Karma based markets, and try to understand when Karma works better than money.

For monetary markets, we calculate the user equilibrium following \cite{wardrop1952correspondence}; when drivers choose between taking the Lincoln Tunnel or the George Washington Bridge route they will minimize their costs.
Each driver experiences three types of costs: for paying the fee if using Lincoln, paying fuel, and delay costs based on their VOT.
We assume an average vehicle consumption of 6.5 l/100km (36 mpg) \footnote{Average US fleet fuel consumption in 2021, according to US transportation secretary Pete Buttigieg. https://edition.cnn.com/2022/04/01/energy/fuel-economy-rules/index.html.}, and a fuel price of 0.96 \$/l \footnote{Average US fuel price. https://tradingeconomics.com/united-states/gasoline-prices.}.

For Karma markets, we assume a pairwise auction between two consumers, where the highest bid is paid to the society, and no Karma redistribution to take place. 
The bidder wins the auction, if his auction is above a certain Karma threshold price and the highest bid.  
The consumer population is initiated with an average budget of 10 Karma points. 
The costs are modelled similar to the monetary markets.

\begin{figure*} [!ht!]
    \centering
    \includegraphics[width=0.49\linewidth]{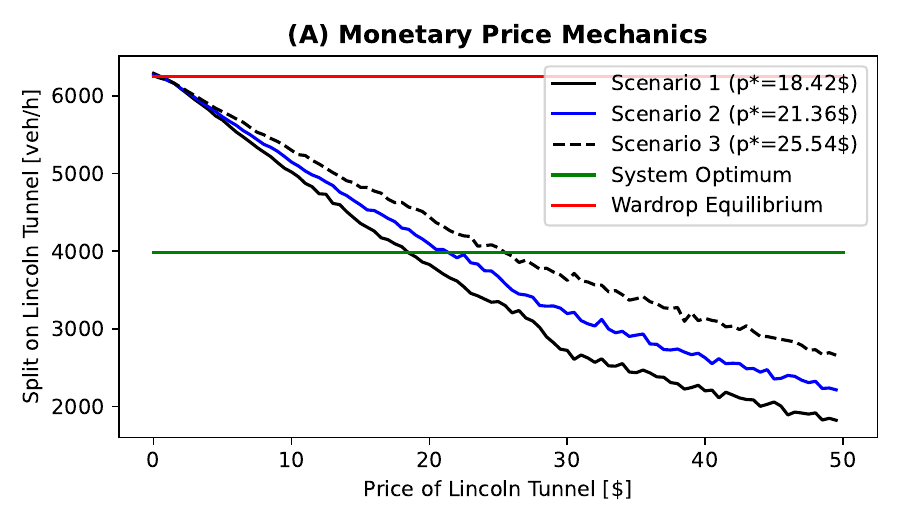}
    \includegraphics[width=0.49\linewidth]{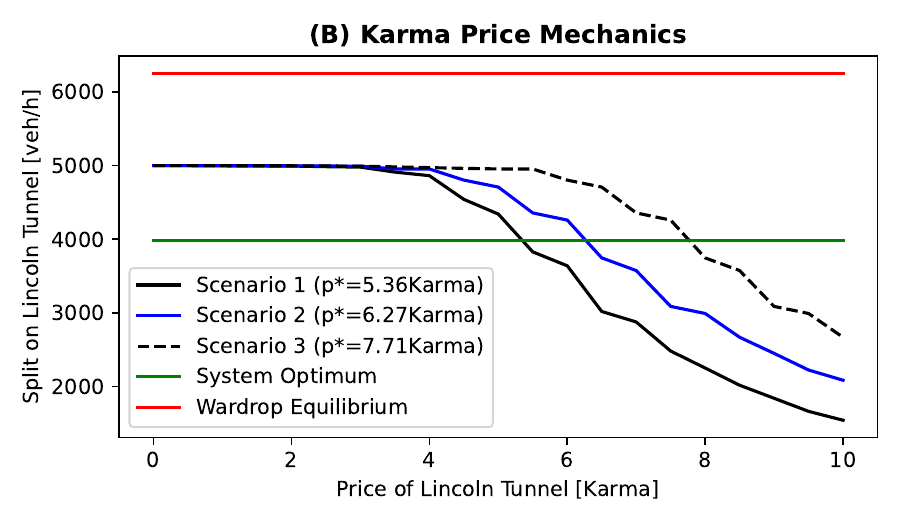}
    \includegraphics[width=0.98\linewidth]{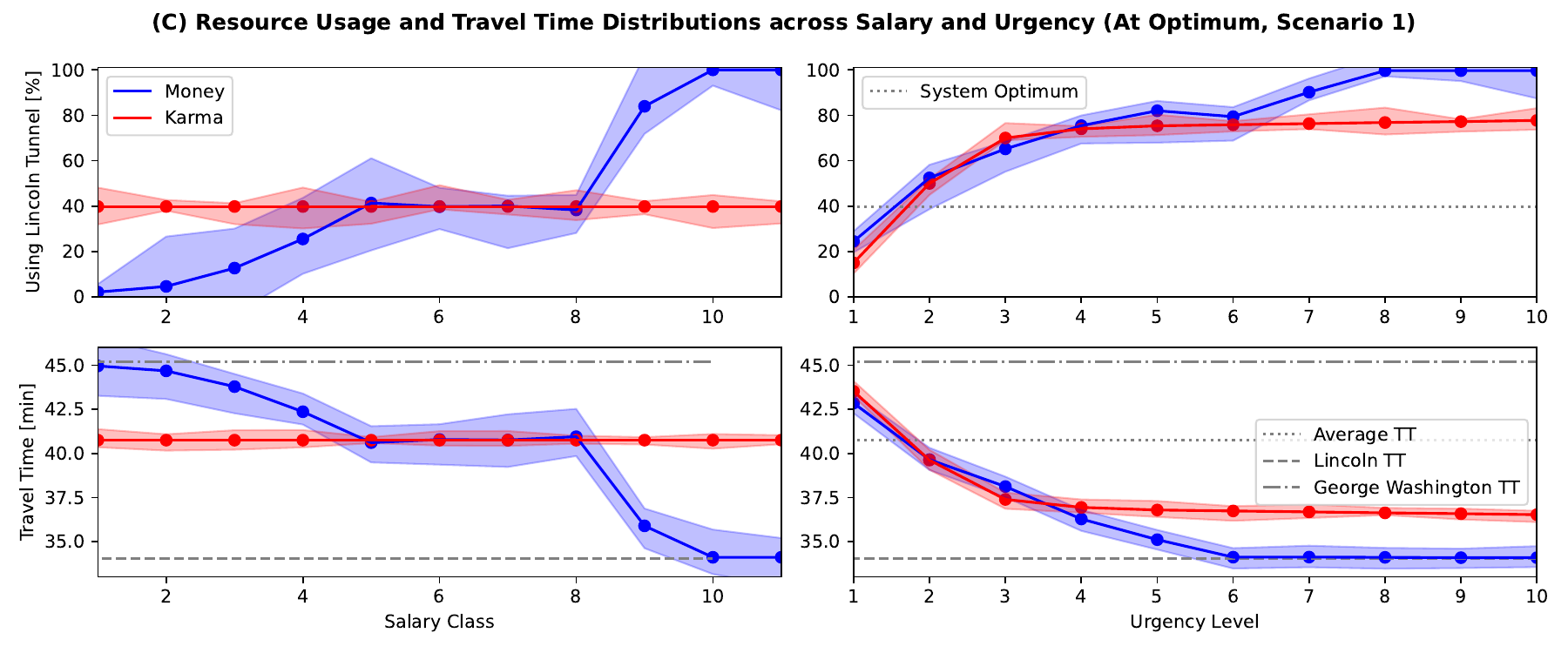}
    \includegraphics[width=0.98\linewidth]{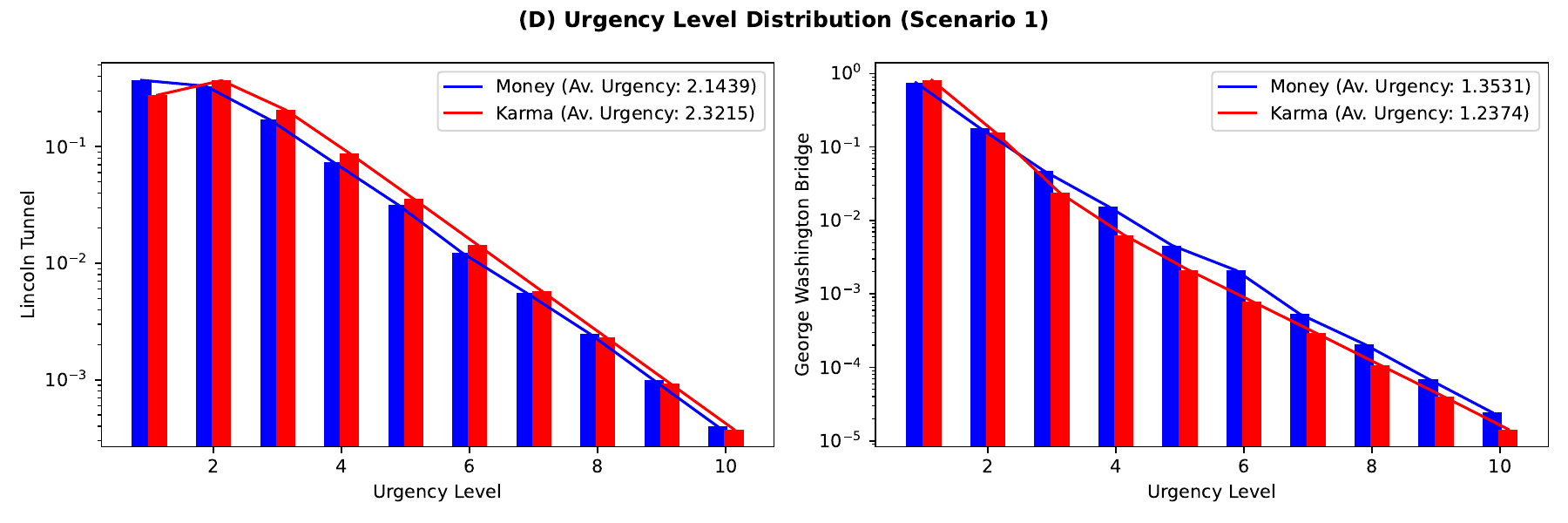}
    \caption{
        \textbf{Case Study: Resource Allocation with Monetary vs. Karma Market} \\ {\scriptsize }
        A systematic comparison of the distributional effects when using money and Karma road pricing of the Lincoln tunnel. \textbf{(A)} Increasing monetary prices incentivizes the population of rational individuals to transition from the Wardrop Equilibrium towards the system optimum for a price of around \$ 20-30. \textbf{(B)} Increasing minimum bids (prices) in Karma auctions leads to the migration of the stationary Nash Equilibrium from 50\% using the tunnel route, to a system-optimal traffic flow split for a threshold between 5 and 8 Karma points. \textbf{(C)} When optimal-pricing the tunnel route with monetary mechanisms, individuals with higher incomes (salary class) experience significantly shorter travel times, as they can afford to access the faster route more often. When pricing with Karma mechanisms, the access to the faster route and travel times are not related to the individuals' income. With regards to the urgency levels, for both mechanisms it can be observed, that individuals with higher urgency experience shorter travel times. In monetary mechanisms, individuals of very high urgencies (beginning from level 6) almost all access the faster route, while in Karma mechanisms, this is not the case. \textbf{(D)} The urgency level distribution of individuals (conditional probability) on both routes reveals, that Karma achieves a greater alignment with the needs, than money does (higher resp. lower average urgency level on faster resp. slower route).
        \\
    }
    \label{fig:casestudy_congestionpricing}
\end{figure*}

Fig.~\ref{fig:casestudy_congestionpricing}(A) shows how different prices for the Lincoln Tunnel will affect the consumer behaviour in monetary markets.
Without the presence of pricing, the user equilibrium lies at the Wardrop equilibrium (around 60\% will take the Lincoln tunnel).
For an increasing price, the demand drops. For a price of \$ 18.42 the user equilibrium lies exactly at the system optimum in scenario 1 (\$ 21.36 and \$ 25.54 for the other two scenarios).
Similarly, a price in Karma Markets can reduce consumption, as shown in Fig.~\ref{fig:casestudy_congestionpricing}(B). 
Without the presence of pricing, the highest-bid auctions with exactly two players will result in always 50\% of the population to win.
For a price of around 5.36 Karma, the user equilibrium can be controlled to sustainable levels as well in scenario 1 (6.27 Karma and 7.71 Karma for the other two scenarios).

Fig.~\ref{fig:casestudy_congestionpricing}(C) shows the share of consumers and travel times across different incomes and urgencies at the optimal price for scenario 1. 
Monetary markets enable consumers with higher incomes (higher salary levels) to use the Lincoln tunnel, and to thus achieve significantly lower travel times. For instance, only $\sim$20\% of the consumers at the lower end of income are willing (or able) to pay for using the Lincoln Tunnel, while the consumers at the upper end of income almost all ($\sim$95\%) are willing to pay for usage. Therefore, consumers with lower income have a noticeably higher travel time ($\sim$45 min) when compared with those of higher incomes ($\sim$34 min).
These results exemplify the equity issues related to road pricing using monetary markets.
Even though the pricing mechanism allows for achieving more efficient usage of the road infrastructure, it embodies the discrimination based on income, which is unevenly distributed across consumers.
Contrary to that, we can observe that Karma markets are completely indifferent to the income of consumers, as Karma follows its own, non-monetary logic. In Karma markets, 39.83\% of all consumers, indifferent from the salary, will get access to the Lincoln tunnel, and therefore achieve on average 40.75 minutes.

With regard to different levels of urgency, we find that monetary markets guarantee access to the resource of interest to the highest levels of urgency (almost 100\%), while only $\sim$22\% of consumers of lower urgency use the tunnel.
Compared to money, the Karma mechanism deviates slightly; in the lowest urgency level 1 $\sim$17\% (so $\sim$5\% less) of the consumers get access to the tunnel, and at urgency level 3 $\sim$70\% (so $\sim$5\% more) of the consumers can access the tunnel. 
For higher levels of urgency, the Karma mechanism only guarantees $\sim$75\% (25\% less) of the consumers access to the tunnel.
At first sight, these results might imply, that Karma does not align as well to the urgencies of consumers, as money does. 
However, one must take into account, that most consumers are in the lower urgency regimes (e.g.\ 95\% of consumers with urgency less than level 5).

Therefore, we have analysed the distribution of urgency levels within the preferred route (Lincoln tunnel) and the alternative route (George Washington bridge), as shown in Fig.~\ref{fig:casestudy_congestionpricing}(D). 
The results show, that the Karma mechanism leads to a situation, where consumers of higher urgency are present in the Lincoln tunnel, and consumers of lower urgency are present on the George Washington bridge, when compared with the monetary market. The Karma mechanism achieves an average urgency level of 2.32 in the Lincoln tunnel (2.14 for money), and an average urgency level of 1.24 on the George Washington bridge (1.35 for money).

Next, we analysed the costs and benefits of the road pricing strategy (for scenario 1).
Table~\ref{table_cba} contrasts resource usage, travel times, and cost breakdown for a situation where there is no pricing, where pricing with monetary markets is applied, and where pricing with Karma markets is applied.
Due to the introduction of pricing, resource usage can be reduced and therefore total travel time (and average travel time) can be reduced to a possible minimum.
The cost breakdown reveals, that the total financial costs per user, that consist of fuel costs, fees for the usage of the Lincoln tunnel, and travel time costs (due to the VOT), can be reduced from \$89.03 (unpriced) down to \$81.20 (9\% less).
Thus, controlling the access to the resource yields significant improvements for the drivers.
While the fuel costs increase only slightly from \$2.19 up to \$2.53 (as more consumers drive the longer route via the George Washington bridge), the pricing introduces financial fee costs in case of the monetary markets of \$7.33 on average to the consumer.
Karma has an advantage here, as no additional financial costs due to fees are generated.
With regards to the costs due to travel times (VOT), we can observe that monetary markets can achieve stronger cost reductions from \$86.84 down to \$71.97 (17\% less), as the money mechanism takes salaries and VOTs into account. Au contraire, Karma focuses on consumer needs (urgencies) only, and hence achieves solely 9\% travel time cost reductions.
Essentially, Karma and money both yield improvements when compared to the unpriced situation, with almost similar total cost improvements per user. The cost reductions from the monetary mechanism originate from a better alignment to the VOTs, at the cost of an additional fee, while the cost reductions from the Karma mechanism originate from a better alignment to the urgencies.

\begin{figure} [ht!]
    \centering
    \includegraphics[width=\linewidth]{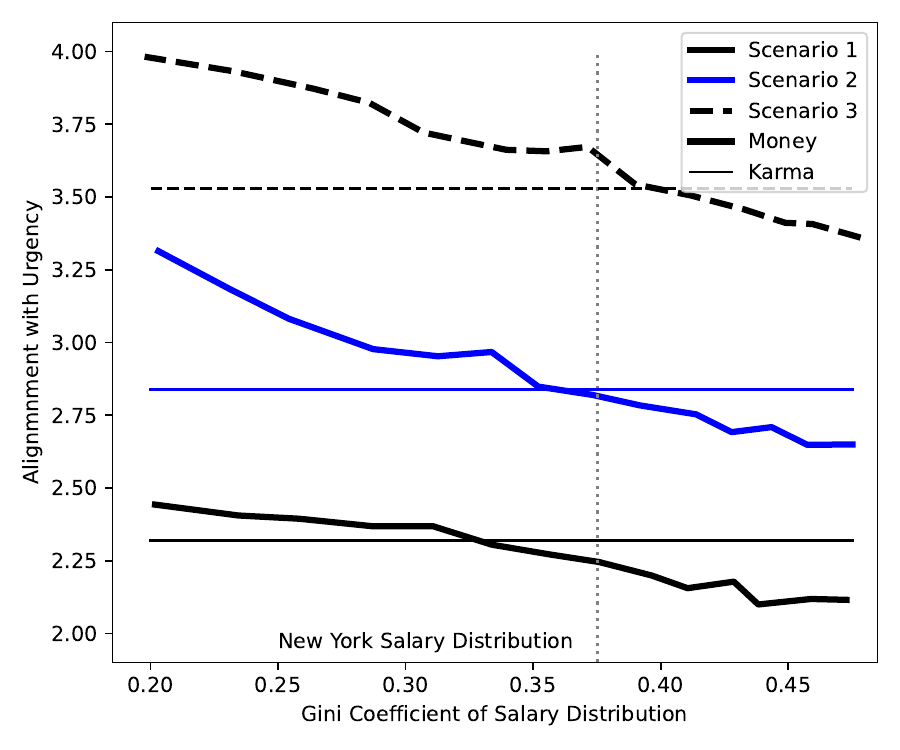}
    \caption{
        \textbf{When Karma Works Better} \\ {\scriptsize }
        The superiority of Karma over money in terms of needs-alignment depends on the distribution of salaries. For a certain level of inequality in the income distribution, Karma performs better than money (Gini coefficient between 0.35 and 0.38). 
    }
    \label{fig:result2}
\end{figure}

Finally, we have tried to better understand when Karma outperforms money.
A major determinant of the distributional effects of the monetary mechanism is the distribution of financial power.
Therefore, we sampled synthetic salary distributions with different levels of evenness measured by the Gini coefficient.
The Gini coefficient assumed salary distribution of New York from the case study lies around 0.375.
We generated salary distributions with Gini coefficients between 0.20 and 0.50, as most nations possess income distributions in that range.
We then determined optimal prices to achieve system optimal resource usage of the Lincoln tunnel, and quantified the average urgency levels of consumers in the Lincoln tunnel as a measure for how well the mechanism allocates resources and how strongly it is aligned to the consumer needs.
The results for the three different scenarios are shown in Fig.~\ref{fig:result2}, where alignment with needs (urgency) is measured as the average urgency level in the Lincoln tunnel.

In societies with more even distributions (smaller Gini coefficients), monetary markets can achieve a larger alignment to the consumer needs.
The larger the inequalities in financial power become, the less alignment with consumer needs can be achieved.
The Karma mechanism instead does not react sensitive to the salary distribution. 
The superiority of Karma over money depends on the urgency distribution as well.
In our case study, it turns out that the Karma mechanism works better than money for scenarios 1 and 2.
In the fictional scenario 3 however, when urgency regimes become more evenly distributed, the Karma mechanism is slightly worse.
In scenario 1, less than 5\% of the consumer population is urgent enough to be willing to pay more than three times hourly salary in exchange for the same amount of time, while in scenario 3 it is already 30\%, which can be considered to occur rarely in practice.

\begin{table}[H]
    \centering
    \begin{tabular}{lrrr}
        \hline
        \textbf{Comparison}              & \textbf{Unpriced} & \textbf{Money} & \textbf{Karma} \\
        \hline
        \textbf{Resource Usage}          &                   &                &                \\
        ...Consumers on Lincoln Route      & 61.79\%           & 39.83\%         & 39.83\%         \\
        \hline
        \textbf{Times}                   &                   &                &   \\
        ...Av. Travel Time {[}min{]} & 45.01             & 40.75          & 40.75          \\
        ...TotalTravelTime {[}veh x h{]}   & 7511              & 6791           & 6791           \\
        \hline
        \textbf{Costs per User}          &                   &                &                \\
        ...Fuel                          & \$2.19            & \$2.53    & \$2.53    \\
        ...Fees                          & \$0.0             & \$7.33       & \$0.0              \\
        ...Travel Time                   & \$86.84           & \$71.97    & \$78.66    \\
        \hline
        \textbf{Total Costs per User}            & \textbf{\$89.03}           & \textbf{\$81.84}    & \textbf{\$81.20}   \\
        \hline
    \end{tabular}
    \caption{\textbf{Case Study: Cost \& Benefit Analysis of Karma Pricing}}
    \label{table_cba}
\end{table}


To summarize, Karma is a fair and efficient resource allocation mechanism that has the potential to address the equity issues of economic instruments when coping with public goods.
The results of the case study indicate, that similar to money, Karma can be used as a resource pricing mechanism, to control the user equilibrium to sustainable levels of consumption.
Contrary to money, Karma embodies fairness, as it does not discriminate based on financial power (income), but orients resource allocation on the urgency of consumers.
Karma is an efficient and robust resource allocation mechanism, that works independently of the distribution of financial power in a society.
The results indicate, that Karma achieves a resource allocation that is better aligned with the urgencies of consumers than money does.
This is especially the case for societies with higher inequalities in financial power.
Furthermore, Karma pricing does not impose additional costs to the users, and generates significant improvements both in travel times and total costs per user.

\section{Conclusions} \label{sec:conclusions}

This study set out to demonstrate the potential of Karma to address equity issues of economic instruments when coping with public goods.
Within this work, we reviewed important design elements and challenges when designing Karma mechanisms, discussed the value proposition of Karma as a non-monetary resource allocation mechanism, model Karma as a game, and equip the reader with a useful software framework to predict consumer behaviour in Karma economies.
At the example of a case study on bridge tolling, we systematically compare the distributional effects and related equity issues of monetary and Karma resource allocation mechanisms and demonstrate the potential of this concept.
The results show, that Karma outperforms money, especially in situations where financial power is unequally distributed across the population, as Karma achieves a stronger alignment to consumer needs.
The Karma mechanism embodies fairness, as it does not discriminate based on financial power (income), and furthermore it does not generate an additional financial burden to the consumers.

Future research could investigate the effects of different urgency distributions, and populations with heterogeneous temporal preferences (discount rates).
Moreover, empirical studies could be conducted to determine more realistic distributions of urgencies and temporal preferences across populations.


\subsection*{CRediT authorship contribution statement}
\textbf{Kevin Riehl: } Conceptualization, Methodology, Software, Formal analysis, Investigation, Data Curation, Writing - Original Draft, Visualization, Project administration. 
\textbf{Anastasios Kouvelas: } Writing - Review \& Editing, Validation.
\textbf{Michail Makridis: } Writing - Review \& Editing, Validation, Supervision.

\subsection*{Declaration of competing interest}
None.

\subsection*{Data availability}
The software framework and documentation can be found on the GitHub repository.

\href{https://github.com/DerKevinRiehl/karma_game_library}{https://github.com/DerKevinRiehl/karma\_game\_library}

\subsection*{Acknowledgements}
We would like to thank Ezzat Elokda, Florian Dörfler, and Saverio Bolognani for the valuable and useful feedback and comments, which was instrumental when improving this work.
We would like to thank Dario Morandini for reviewing and testing the software framework, and his helpful feedback.

\clearpage
\renewcommand*{\bibfont}{\normalfont\small}
\printbibliography

\clearpage
\appendix
\renewcommand\thefigure{A\arabic{figure}}
\setcounter{figure}{0}

\section{Assumptions of Karma Game}
\label{sec:game-assumptions}
Karma is a \textbf{repeated, dynamic population game}, as the Karma game is not played once but multiple times and the time is split into discrete epochs (rounds). 
As argued in previous works, a time horizon of at least multiple rounds is crucial to incentivize cooperation amongst selfish participants~\cite{article_185,article_600,article_1005,article_college_A,article_college_B}.
The formalism describes a \textbf{stochastic game}, as the state and state transitions of agents depend upon probabilities. 
Besides, the behaviour of agents, such as their bids or accepting resource provision requests, is modelled as a Markov decision process~\cite{article_600,article_1001}.
The formalism describes a \textbf{population game}, as the Karma mechanism aims to represent the strategic interplay in large societies of rational (selfish) agents (participants)~\cite{article_1001}.

The Karma game makes certain assumptions in order to facilitate simulation and computation of the stationary Nash equilibrium.
In this chapter, we will discuss these assumptions, as  well as provide guidance on how intelligent modelling can achieve complex Karma games despite these restricting assumptions: 

\begin{itemize}
    \setlength\itemsep{0em}
	\item The selection of participants for an interaction is randomly chosen from the population.
	\item The Karma balance of each agent must be greater or equal to zero. There is no such thing as a Karma debt allowed.
	\item The total amount of Karma in circulation remains constant over time.
	\item The total amount of agents in the population remains constant over time.
	\item The possible actions of an agent in an interaction solemnly depends on its Karma balance.
	\item The decision making process of agents to choose an action during an interaction solemnly depends on its own, private state (type, urgency, Karma), and not the states of others.
	\item The decision making process of agents to choose an action during an interaction solemnly depends on its own, current state, and not on its previous states. 
	\item The agents have an identical decision making process by following the same, state-specific policy, assuming that the policy describes the best possible choice (optimum) for an egoistic, selfish, rationally acting agent.
\end{itemize}

Important implications of these assumptions, guidance on modelling, and how intelligent modelling can achieve complex Karma games despite the restricting assumptions is discussed in the following.

\begin{itemize}
    \setlength\itemsep{0em}

	\item The selection of participants is not discriminatory in terms of agent type, urgency or Karma balance. This means, that the probability to be selected in an interaction is proportional to the share in the distribution defined by $d$. There must be at least two participants selected. In certain contexts, it could be possible to more than two participants, or even the whole population as in \cite{article_1002}.
	
	\item A negative Karma balance is not possible. The set of possible Karma balances however could be defined as infinite, choosing the positive natural numbers.

	\item A changing number of agents could be modelled by introducing a specific type and urgency, for agents that have no cost and no temporal preference, and thus will act in an interaction with the specific action type ``no action'' (e.g.\ refusing any action), so that they always result in a specific outcome type (not receive resource).

	\item Only sealed bid auctions as a form of interaction are possible in the Karma system, as the state of agents is private to others and the system. 
    This is particularly important, as it enables this form of decentralized, parsimonious control to work without a large overhead, and without the need to exchange any information except for the action. Of course this assumes that certain forms of smart contracts enforce all participants to act according to the set of possible actions.

	\item An important assumption this model makes is, that the agent's decision making is a Markov decision process chain, meaning that the decision making of the agent at a specific stage only depends upon the agent's current exogenous state and available actions, but not a history on past states or actions~\cite{article_348}. 

	\item The action of an agent must at least be the bid in a sealed bid auction, but can in addition include other actions. As an example, in~\cite{article_1002} the action is the bid and the decision when (timeslot) to start driving to work in the morning.

	\item The outcome of the interaction $o_e$ is the vector of outcomes $o_{e,j}$ for each participant $j$. It could be modelled binary, where $o_{e,j}=1$ represents that participant $j$ receives the resource, and $o_{e,j}=0$ represents that participant $j$ does not receive the resource.

	\item Multi-agent control systems aim to find mechanisms to align selfish behaviour with global, societal goals. One can assume, that a population of rational agents, be it human or autonomous, will always try to express a behaviour that will maximize their own, expected reward. Therefore, once proven that the policy is the optimal policy for a rational, selfish agents, we can conclude that all agents will follow the identical, state specific policy. Of course, this does not necessarily mean that each agent has the (cognitive) ability to identify the optimal policy, but this can be discussed in future research. Besides, one could either assume an algorithmic bidding assistant for humans.

\end{itemize}

The interested reader is recommended to look in further modelling peculiarities in~\cite{article_567,article_600,article_1002,elokda2023dynamic}.

\section{Stationary Nash-Equilibrium Computation}
\label{sec:optimizers}

In order to calculate the optimal social state at the stationary Nash equilibrium, previous works used optimization algorithms that are iterative, heuristic and numeric. 
\cite{article_567} suggests fixed point computation, momentum method and simulated annealing. 
\cite{article_600,article_1002,article_1004} use evolutionary dynamics inspired optimization algorithms. 
\cite{article_1005} employs the Smith protocol. 
The choice of a suitable optimization algorithm is important, as the state-space is large and the dynamics are rigid~\cite{article_1005}. 
In this work, we compute the stationary Nash equilibrium employing an evolutionary dynamics inspired optimization algorithm~\cite{article_600}.
The interested reader is highly recommended to review further optimization approaches in~\cite{sandholm2010population}, such as the Replicator-approach, Brown-von-Neumann-Nash, Smith, and the Projection.



\end{document}